 \documentclass[final,3p,times]{elsarticle}

\usepackage[T1]{fontenc}
\usepackage[utf8]{inputenc}

\usepackage{amssymb}
\usepackage{amsmath}
\usepackage{hyperref}

\usepackage{ar}
\usepackage{booktabs}
\usepackage{color}

\biboptions{sort&compress}

\usepackage[final]{changes}

\journal{Journal of Fluid and Structures}

\begin{document}

\begin{frontmatter}



\title{Constructive interference in a network of elastically-bounded flapping plates}  


\author[dicca,infn]{S. Olivieri\corref{cor1}}   \ead{stefano.olivieri@edu.unige.it}
\author[difi]{C. Boragno}
\author[dii]{R. Verzicco}
\author[dicca,infn]{A. Mazzino}

\address[dicca]{DICCA (Dipartimento di Ingegneria Civile, Chimica e Ambientale), University of Genoa, Via Montallegro 1, I-16145, Genoa, Italy}
\address[infn]{INFN (Istituto Nazionale di Fisica Nucleare), Genoa Section, Via Dodecaneso 33, I-16146, Genoa, Italy}
\address[difi]{DIFI (Dipartimento di Fisica), University of Genoa, Via Dodecaneso 33, I-16146, Genoa, Italy}
\address[dii]{DII (Dipartimento di Ingegneria Industriale), University of Rome `Tor Vergata', Via del Politecnico, I-00133, Rome, Italy}

\cortext[cor1]{Corresponding author.}

\begin{abstract}
Aeroelastic phenomena are gaining significant attention from the perspective of energy harvesting (EH) with promising applications in supplying low-power remote sensors. 
Besides the development of individual EH devices,
further issues are posed when considering multiple objects for realizing arrays of devices and magnifying the extracted power. Due to nonlinear mutual interactions, the resulting dynamics is generally different from that of single devices and the setup optimisation turns out to be nontrivial.
In this work, we investigate the problem focusing on a flutter-based EH system consisting of a rigid plate anchored by elastic elements and invested by a uniform laminar flow, undergoing regular limit-cycle oscillations and flapping motions of finite amplitude.
We consider a simplified, yet general, physical model and employ three-dimensional direct numerical simulations based on a finite-difference Navier-Stokes solver combined with a moving-least-squares immersed boundary method.
Focusing on main kinematic and performance-related quantities, we first report on the dynamics of the single device and then on multiple devices, considering different arrangements (i.e.: in-line, staggered and side-by-side). A parametric exploration is performed by varying the mutual distance between the devices and insights are provided. For the in-line arrangement, a recovery in performance for downstream devices is achieved by tuning their elasticity. Moreover, cooperative effects in the side-by-side arrangement are found to be substantially beneficial in terms of resulting power, with increases (i.e. constructive interference) up to 100\% with respect to the single-device configuration.
In order to confirm this numerical evidence, complementary results from wind-tunnel experiments are presented.
Finally, we describe the system behaviour when increasing further the number of devices, outlining the ultimate goal of developing a high-performance EH network of numerous aeroelastic energy harvesters.
\end{abstract}

\begin{keyword}
flutter \sep flapping  \sep energy harvesting \sep multiple \sep in-line \sep side-by-side



\end{keyword}

\end{frontmatter}



\section{Introduction}
\label{sec:intro}

Flow-induced vibrations are receiving growing interest from the perspective of energy harvesting (EH), in view of applications related to the supply of low-power sensor networks within the framework of the Internet of Things~\cite{li2016review,mccarthy2016review}.
A variety of novel devices and corresponding models has been proposed in literature and they can be classified depending on the aeroelastic instability occurring, such as, e.g., flutter, galloping and vortex-induced vibrations. 
In all cases, the understanding of these nonlinear phenomena is crucial in order to design efficient and reliable EH devices and to fine-tune their performance.

For slender aerodynamic bodies such as wings or plates, the typical mechanism to be exploited is aeroelastic flutter, consisting of a self-excitation between flow and structural response (see, e.g., Ref.~\cite{mccarthy2016review} for an introduction to flutter-based EH).
Among the various flutter-based EH systems, one can include those based on flapping flags~\cite{tang2009cantilevered,michelin_doare_2013,shoele_mittal_2016} as well as passively flapping airfoils~\cite{xiao2014review,young2014review}. For the latter, the motion of an essentially rigid, streamlined body is limited to a certain number of degrees of freedom (DoFs), typically two: translation along the transverse direction (plunge) and rotation around a spanwise axis (pitch).

Focusing on fully-passive systems (i.e., those whose dynamics is entirely governed by the fluid-structure interaction, without any prescribed kinematics), early research by Peng and Zhu~\cite{peng2009}, conducted by two-dimensional numerical simulations, highlighted the variety of possible flapping states and provided a first estimate of the resulting power and efficiency.
These findings were confirmed and enriched by further computational studies in closely comparable conditions~\cite{zhu2012shear,wang2017structural} as well as rather different systems, e.g. with different cross sectional shapes and/or operating at higher-Reynolds regimes~\cite{young2013numerical,veilleux2017numerical,ramesh2015intermittent,wang2018lco}.

Young et al.~\cite{young2013numerical} introduced a configuration where the pitch and plunge motions are constrained by a mechanical linkage along with pitch control in order to increase the performance.
Veilleux and Dumas~\cite{veilleux2017numerical} performed an optimization study for a fully-passive device by two-dimensional CFD simulations which became the basis for the experimental prototype later presented by Boudreau et al.~\cite{boudreau2018}. These studies agree in reporting that the performance is improved for an adequate synchronization between the two DoFs and in case of  nonsinusoidal pitching motion.
Wind-tunnel investigations were reported by Pigolotti et al.~\cite{pigolotti2017destabilizing,pigolotti2017jsv,pigolotti2017jfs} considering a flat plate in a classical pitch-and-plunge arrangement and exploring systematically the effect of several physical parameters on the flutter onset and the nonlinear oscillations in the postcritical regime. 
A nearly identical system was considered in the work by Wang et al.~\cite{wang2018lco} where two-dimensional computations were performed in order to characterize the dependence of limit-cycle oscillations with respect to the governing parameters, initial conditions, spring nonlinearity and extraction (modelled by viscous damping).

Besides the development of the individual energy harvester and the consequent understanding of the associated nonlinear system, 
a further degree of complexity arises when looking at a network of multiple energy harvesters, that allow to scale the total extracted power. 
In fact, the behaviour of each device of the network is coupled to all the others due to the nonlinear character of the dynamical system.

Among the few contributions given in this latter perspective, Bryant et al.~\cite{bryant2012wake} tested devices made by a rigid airfoil hinged on a cantilevered flexible beam and arranged in in-line or staggered configuration, reporting performance improvements for downstream devices due to the beneficial effect of wake forcing.
Moreover, it was later proposed that such effect can be controlled by tuning the pitching stiffness of downstream devices~\cite{kirschmeier2018wake}.
McCarthy et al.~\cite{mccarthy2013downstream,mccarthy2014visualisation} focused on the in-line arrangement considering a different flutter-based system where a triangular leaf is joined to a piezoelectric stalk by a revolute hinge. They reported increases in power of about $40\%$ for the downstream device caused by the interaction with the horseshoe cone vortex released by the upstream device.

Even if they have a different behaviour, flapping flags were also investigated in the same framework. Several works focused, in particular, on multiple filaments or flexible plates placed in the side-by-side configuration, showing increased oscillations over an intermediate range of distances between the flags~\cite{zhang2000flexible,favier2015,huertas-cerdeira2018}.

In this work, we focus on an aeroelastic system based on elastically-anchored plates experiencing a fluttering instability when invested by laminar flow, giving rise to finite-amplitude limit cycle oscillations (LCOs) and flapping motions~\cite{boragno2012elastically,orchini2013flapping,olivieri2017fluttering,olivieri2017aeroelastic,boccalero2017power}.
Previous related work addressed the identification of the critical threshold for sustained flapping~\cite{orchini2013flapping,olivieri2017fluttering} and a first characterization of the postcritical behaviour for a representative configuration of the real device~\cite{olivieri2017aeroelastic}, as well as the development of an experimental prototype equipped for energy extraction by electromagnetic coupling~\cite{boccalero2017power}.
Overall, we aim at developing EH devices of centimetric size
able to
extract electrical power $\mathcal{O}(\mathrm{mW})$ in low wind conditions (i.e., $U < 5 \, \mathrm{m/s}$).
Despite the analogies with flutter-based systems, 
differences exist with respect to the classical pitch-and-plunge model, the main being the absence of rotational elastic constraints yielding a different dynamics, e.g. in the critical condition for flapping onset~\cite{olivieri2017fluttering}.

The goal of the present work is to provide a fundamental study on the behaviour of multiple elastically-bounded flapping plates immersed in an incompressible laminar flow, 
characterizing the resulting dynamics and giving useful insights for the development of EH networks made of arrays of such devices.
To this aim, we will employ three-dimensional numerical simulations combining a finite-difference Navier-Stokes solver and an immersed boundary technique. First we will characterize the dynamics of the single device, and then investigate three basic arrangements of multiple devices, i.e.: (i) in-line, (ii) staggered and (iii) side-by-side. A parametric study will be conducted by focusing on the dependence of main quantities of interest with respect to the mutual distance between devices.
In order to corroborate the numerical evidence, we will provide experimental results from wind-tunnel measurements pertaining to real EH applications.

Following this introduction, the rest of the paper is structured as follows. In Sec.~\ref{sec:physical-model} we introduce the representative aeroelastic system and the main governing parameters; Sec.~\ref{sec:num-method} concerns the numerical method, while results are provided in Sec.~\ref{sec:results}. Conclusions are drawn in Sec.~\ref{sec:conclusions}.
In a final Appendix we provide evidence for the validation of the numerical method and for all the run parameters used for the simulations.

\section{Physical model}
\label{sec:physical-model}

\begin{figure}
\centering
\includegraphics{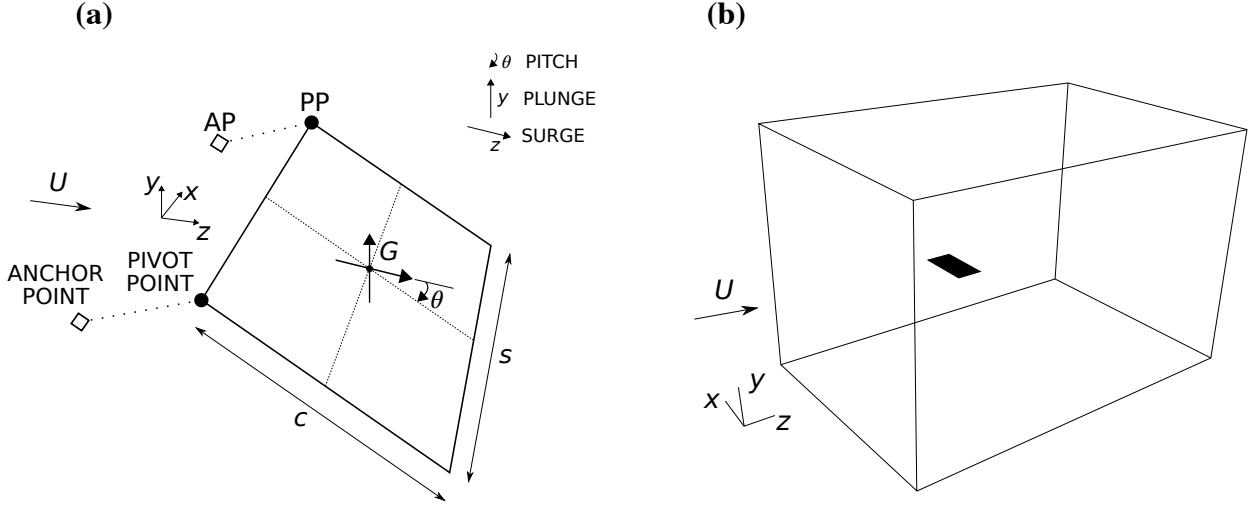}
\caption{Sketches of \added{(a) aeroelastic model considered in the present work} and (b)  domain used for numerical simulations.}
\label{fig:sketch}
\end{figure}

We consider a rigid plate of homogeneous density $\rho_\mathrm{s}$ and geometry characterized by chord
$c$, span $s$ and thickness $\delta$. 
 The plate is restrained by two linear springs with stiffness $k$ and zero restlength.
 As shown in Fig.~\ref{fig:sketch}a, for each spring one end is connected to the plate at the pivot point (PP), moving with the body, while the other one is retained fixed at the anchor point (AP).
The body is invested by a uniform flow with inflow velocity $U$, the fluid has kinematic viscosity $\nu$ and density $\rho_\mathrm{f}$.
As additional assumptions, gravity effects are neglected and
energy extraction is not considered since we are concerned with the purely aeroelastic behaviour of the system.

Looking at the introduced quantities, the following four nondimensional parameters can be derived:
the chord-based Reynolds number $\mathit{Re}=U c / \nu$, the nondimensional stiffness $K = 2\, k /(\rho_\mathrm{f} U^2 c)$ (accounting for both springs), the two-dimensional mass parameter $\rho_\mathrm{w} = \rho_\mathrm{s} \delta / (\rho_\mathrm{f} c)$ \added{(where the subscript `w' stands for wing, this quantity depending both on the density ratio and the wing cross section)} and the planform aspect ratio $\AR = s/c$.
In the rest of the work we will always refer implicitly to nondimensional quantities, including spatial and temporal coordinates, i.e.: $ (x,y,z) \mapsto (x/c,y/c,z/c)$, $t \mapsto t /(c/U)$.

Three degrees of freedom (DOFs) are allowed in the plate motion: (i) translation in the streamwise direction (\textit{surge}), (ii) translation in the transverse direction (\textit{plunge}), (iii) rotation around the spanwise axis (\textit{pitch}).  Under these assumptions, the rigid body equations governing the motion of the center of mass (whose position is denoted by $\mathbf{r}_G$) and rotation $\theta$ (written with respect to $G$) read:
\begin{equation}
m \, \ddot{\mathbf{r}}_G = \mathbf{F}_\mathrm{aero} +  \mathbf{F}_\mathrm{el}, 
\label{eq:rbm1}
\end{equation}
\begin{equation}
I_x \, \ddot{\theta} = {M}_\mathrm{aero} + M_\mathrm{el},
\label{eq:rbm2}
\end{equation}
where $m = \rho_\mathrm{w} \, \AR$ is the mass, $I_x$ is the moment of inertia with respect to an axis passing through the hinge and directed in the spanwise direction, $\mathbf{F}\added{_\mathrm{aero}}$ is the aerodynamic force, $\mathbf{F}_\mathrm{el} = -K\, \mathbf{r}_{PP}$ is the elastic force exerted by the springs, ${M}_\mathrm{aero}$ is the aerodynamic moment and ${M}_\mathrm{el} = (\mathbf{r}_{PP} - \mathbf{r}_G) \times \mathbf{F}_\mathrm{el}$ is the elastic one.
The aerodynamic force and moment are obtained by integrating the fluid stress tensor over the plate surface $S$:

\begin{equation}
 \mathbf{F}_\mathrm{aero} = \int_{S} (\boldsymbol{\tau}- p \mathbf{n}) \,  \mathrm{d}S, 
\label{eq:F_aero}
\end{equation}
\begin{equation}
{M}_\mathrm{aero} = \int_{S} \mathbf{r}\times(\boldsymbol{\tau}- p \mathbf{n}) \,  \mathrm{d}S,
\label{eq:M_aero}
\end{equation}
where $\boldsymbol{\tau}$ is the viscous shear stress, $p$ is the pressure, $\mathbf{n}$ is the unit vector normal to the plate and $\mathbf{r}$ is the distance with respect to the center of mass $G$.

Although differences exist with respect to the real device (e.g., in mass distribution), this kind of description, already considered in~\cite{orchini2013flapping,olivieri2017fluttering} and extended here to the three-dimensional case, is able to reproduce the essential physics that has been observed experimentally and thus matches our goal of understanding, at least qualitatively, the peculiar effects arising when multiple devices are considered.

\section{Numerical method}
\label{sec:num-method}

The flow obeys to the incompressible Navier-Stokes equations which, in nondimensional form, read:
\begin{align}
  \frac{\partial \mathbf{u}}{\partial t} + \mathbf{u} \cdot \nabla \mathbf{u} &= -  \nabla p+ \frac{1}{\mathit{Re}} \nabla^2 \mathbf{u} + \mathbf{f}
   \label{eq:NS1}
,\\
  \nabla \cdot \mathbf{u} &= 0,
 \label{eq:NS2}
\end{align}
where $\mathbf{u}=\mathbf{u}(\mathbf{x},t)$ is the fluid velocity, $p=p(\mathbf{x},t)$ the pressure and $\mathbf{f}=\mathbf{f}(\mathbf{x},t)$ a volumetric forcing.
We consider a three-dimensional domain of size $L_x \times L_y \times L_z$ (Fig.~\ref{fig:sketch}b), with the following boundary conditions:
the fluid velocity is uniform at the inlet, convective boundary conditions are used at the outlet~\cite{ferziger2012computational},
the top and bottom faces are treated as slip (i.e., non-penetrating) walls, while periodicity is assumed at side faces.

Since we aim at dealing with multiple moving objects, we resort to the immersed boundary (IB) technique~\cite{mittal2005_review} and employ the moving-least-square (MLS) method recently proposed by de Tullio and Pascazio~\cite{detullio2016} that here we briefly summarize (for more details, see Refs.~\cite{verzicco1996,detullio2016,spandan2017}).

Eqs.~\eqref{eq:NS1} and~\eqref{eq:NS2} are solved numerically on a Cartesian grid, with the forcing term $\mathbf{f}$ mimicking the presence of solid bodies.
Centered finite differencing is adopted for space discretization using a staggered grid, the overall scheme yielding second-order accuracy. The main iteration loop can be summarized as follows:
First, the nonlinear terms are computed explicitly using a second-order Adams-Bashfort scheme. Next, the diffusive terms are discretized implicitly with a Crank-Nicolson scheme and an intermediate, nonsolenoidal velocity field is computed by an approximate factorization technique for the resulting algebraic system~\cite{verzicco1996}.
The correction by the IB forcing is then added to the velocity.
Finally, the Poisson equation for mass conservation is solved and the divergence-free velocity is obtained, together with the pressure field.

Concerning the IB treatment, a moving-least-square (MLS) interpolation is used to reconstruct the solution at the immersed surface~\cite{vanella2009}. 
The plate is discretized by a planar surface mesh with $N_\mathrm{t}$ triangular elements, their centroids being the Lagrangian markers at which the forcing is computed by imposing the no-slip condition.
For each Lagrangian marker, we consider a support domain enclosing $N_\mathrm{e} = 27$ adjacent Eulerian nodes which are used for the interpolation and spreading operations.
Compared to classical IB approaches, the adoption of the described method allows to use a larger simulation timestep and to obtain smoother flow solutions and aerodynamic loads.

The numerical procedure has been tested for the present application by a convergence analysis with respect to the spatial and temporal resolution. Results of this latter and discussion on the choice of grid parameters are collected in~\ref{sec:validation}.
As a baseline indication, the used domain box ranges from $(-5,-5,-5)$ to $(5,5,10)$ and is discretized using a minimum grid spacing $h = 0.02$, with
the triangulated mesh discretizing the solid plate having comparable resolution, so that for the case $\AR=2$ the total number of triangular elements is $N_\mathrm{t} \approx 1.3 \times 10^4$.

\section{Results}
\label{sec:results}

\subsection{Single device}
\label{sec:single}

As a starting point, we characterize the dynamics of the single, isolated device.
Two points are investigated to this aim: (i) the identification of the critical condition for sustained flapping and (ii) the dependence of the limit-cycle features on the governing parameters.
 The first point can be addressed by recalling the predictive arguments of Ref.~\cite{olivieri2017fluttering}, based on the coupling between the natural frequency and the frequency of the pitching response for small angles in the case of hinged plate. Recasting these arguments for our model, the following expression is found for the critical $K$:
\begin{equation}
K_\mathrm{cr} \approx \frac{3}{4} \pi \AR.
\label{eq:K_cr}
\end{equation}

\begin{figure}
\centering
\includegraphics{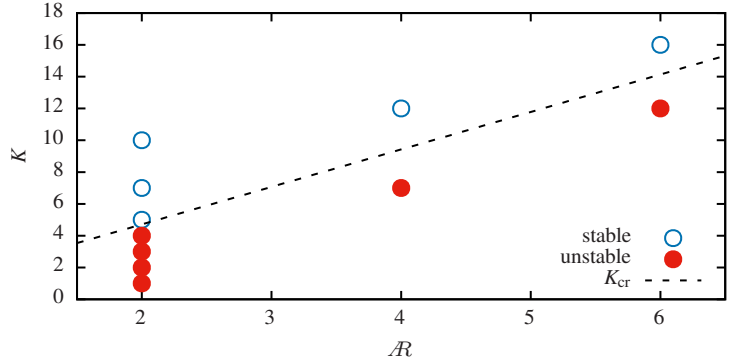}
\caption{
Threshold for sustained flapping in the $(\AR,K)$ plane. Empty circles: stable cases; filled circles: unstable cases. The dashed line reports the theoretical prediction by Eq.~\eqref{eq:K_cr}.
}
\label{fig:K_vs_AR}
\end{figure}

In order to verify this prediction, we have performed a series of simulations at $\rho_\mathrm{w} = 2$ and $\mathit{Re} = 100$ while varying $\AR$ and $K$, whose results are collected in Fig.~\ref{fig:K_vs_AR}.
The plot shows that indeed sustained flapping occurs for $K<K_\mathrm{cr}$ (unstable cases); on the other hand, for $K>K_\mathrm{cr}$ the wing asymptotically aligns with the flow (stable cases). 
As predicted by the theory, a linear dependence of $K_\mathrm{cr}$ on $\AR$ is found.
Although here the simulations have been performed at $\mathit{Re}=100$, the same evidence was found for different values (tested up to $\mathit{Re}=1000$), in agreement with Eq.~\eqref{eq:K_cr} where the Reynolds number does not explicitly appear.
Furthermore, similar behaviour has been verified when varying the density parameter $\rho_\mathrm{w}$.

\begin{figure}
\centering
\includegraphics[width=0.9\textwidth]{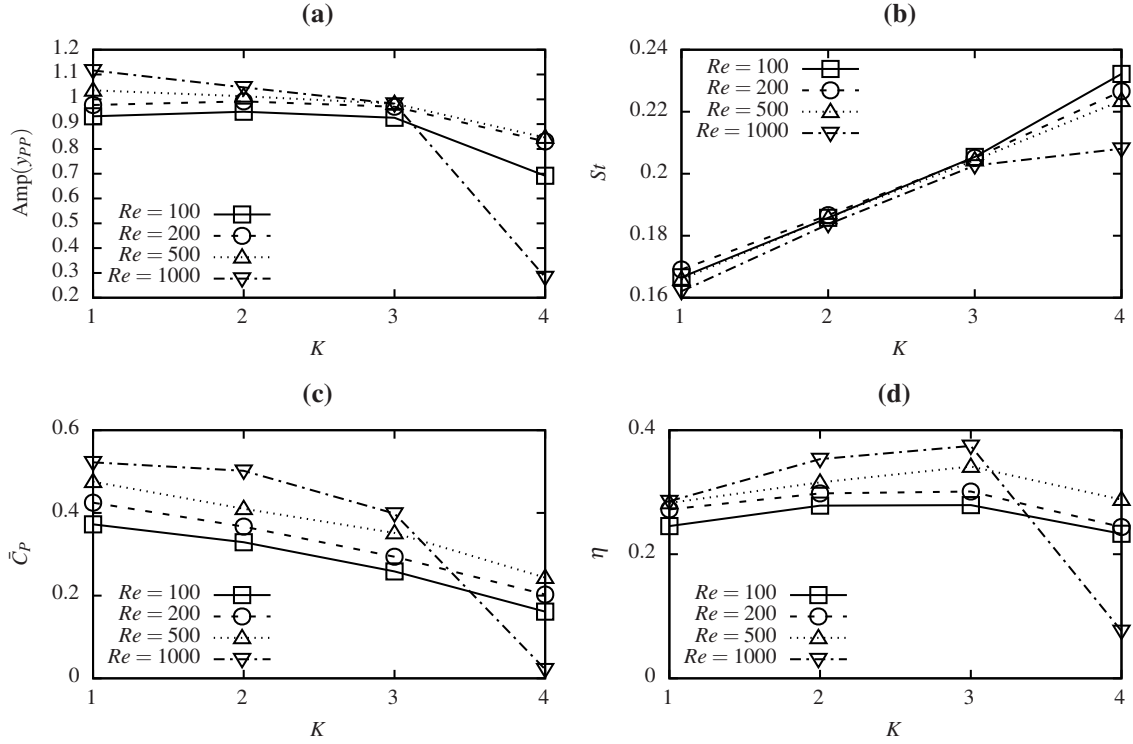}
\caption{
Flapping observables for the single device as a function of $K$ for $\mathit{Re}=\{100$ (squares, solid line), 200 (circles, dashed line), 500 (triangles, dotted line) and 1000 (reverse triangles, dash-dotted line)$\}$,  $\rho_\mathrm{w}=2$ and $\AR=2$: (a) PP transverse oscillation amplitude, (b) Strouhal number, (c)~power coefficient and (d) Betz efficiency.   
}
\label{fig:single}
\end{figure}

Let us now move on to the analysis of the post-critical, nonlinear regime, fixing the aspect ratio to $\AR=2$ and exploring the dependence of the flapping dynamics for different values of the nondimensional stiffness ($K=\{1,2,3,4\}$) and Reynolds number ($\mathit{Re}=\{100,200,500,1000\}$).

The resulting system kinematics and performance are summarized in Fig.~\ref{fig:single}.
The first quantity of interest (also from the EH viewpoint) is the peak-to-peak amplitude of the PP displacement along the transverse distance.
As shown in Fig.~\ref{fig:single}a, this is found to increase for decreasing $K$ (i.e., softening the spring).
On the other hand, the Strouhal number (defined as $\mathit{St}=f c / U$, where $f$ is the flapping frequency)  increases almost linearly with $K$ (Fig.~\ref{fig:single}b).

Concerning the performance parameters, we refer to the average plunge power coefficient $\bar{C}_P$ and the Betz efficiency $\eta$, defined as:
 
\begin{equation}
\bar{C}_P = \frac{\bar{P}_y}{ \frac{1}{2} \, \rho_\mathrm{f} s c \, U^3},
\label{eq:CP}
\end{equation}

\begin{equation}
\eta = \frac{\bar{P}_y}{  \frac{1}{2} \,\rho_\mathrm{f} s d \, U^3},
\label{eq:etaB}
\end{equation}
where $\bar{P}_y = \frac{1}{T} \int_T F_\mathrm{aero}^y \, \dot{y}_{PP} \, \mathrm{d}t$ is the average power associated with the plunge motion, $F_\mathrm{aero}^y$ is the vertical component of the aerodynamic force (i.e., the lift force), and $d$ is the maximum transverse distance swept by the wing during its motion.
Looking at Fig.~\ref{fig:single}c, the trend of the power coefficient appears to be qualitatively similar to those found for the PP amplitude. Conversely, the Betz efficiency exhibits a trend which is not monotonic (Fig.~\ref{fig:single}d).

Overall, variations of the Reynolds number do not appear to modify the described trend.
\added{For lower $\mathit{Re}$, however, the oscillation amplitude (as well as power and efficiency) weakens, consistently with a more dominant effect of viscosity, except for the stiffest case $K=4$ where for $\mathit{Re}=1000$ we observe a sharp decrease. In this case, it can be observed that the wake remains attached to the wing, while for lower $\mathit{Re}$ a sequence of vortices is released from the leading edge, producing higher oscillations.}

In light of this evidence, in the following we will focus uniquely on the case at $\mathit{Re}=100$. This value is selected to deal with the smoothest flow solution among the considered cases, in order to get a clearer understanding of the basic mechanisms occuring in the interaction between multiple devices. 

\begin{figure}
\centering
\includegraphics{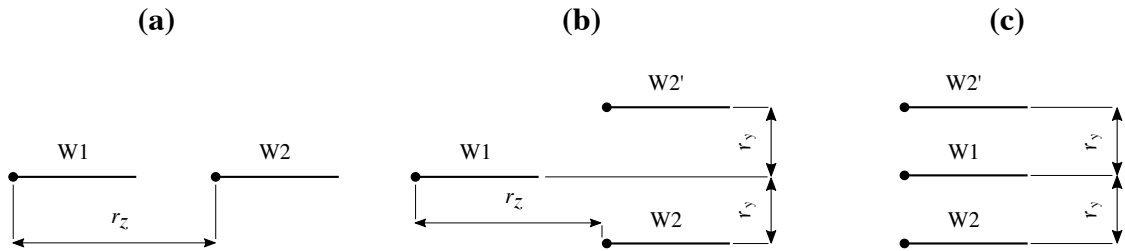}
\caption{   
Configurations investigated for multiple flapping wings: (a) in-line arrangement; (b) staggered arrangement; (c) side-by-side arrangement.
}
\label{fig:sketch-multiple}
\end{figure}

\subsection{In-line arrangement}
\label{sec:in-line}

\begin{figure}
\centering
\includegraphics{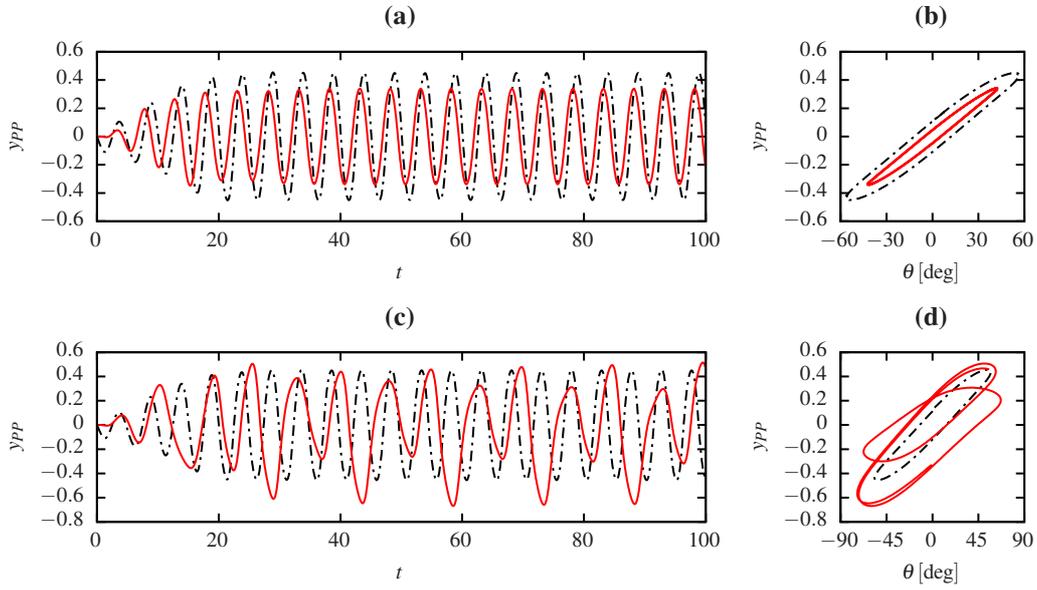}
\caption{Time history of transverse PP oscillation (left panels) and steady-state LCO in ($\theta,y_{PP}$) plane  (right panels) for two devices in-line configuration with $r_z = 2$, (a,b) $K_{(1)}=K_{(2)}=3$ and (c,d) $K_{(1)}=3$, $K_{(2)}=1$. Black dot-dashed line: upstream wing (W1); red solid line: downstream wing (W2).
}
\label{fig:in-line_TH-PS_tuned-K2}
\end{figure}

\begin{figure}
\centering
\includegraphics{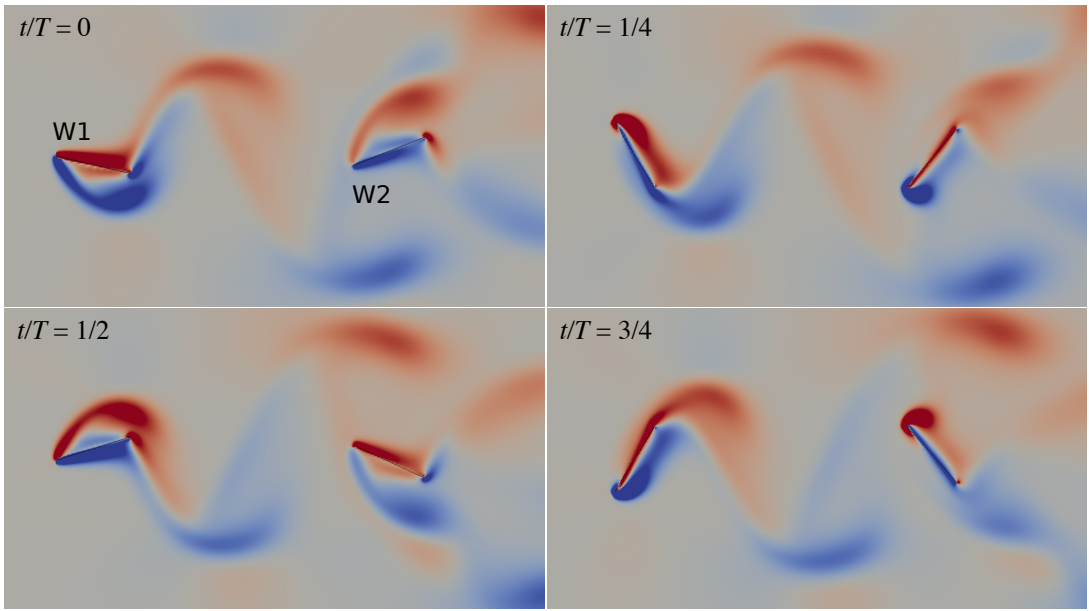}
\caption{Instantaneous views of plate position and vorticity field (negative values (i.e. counterclockwise) in blue, positive ones (i.e. clockwise) in red) within one flapping cycle for the in-line arrangement with $r_z = 4$ and $K_{(1)}=K_{(2)}=3$.
}
\label{fig:in-line_vort}
\end{figure}

We begin our study on the interaction of multiple devices by considering an in-line configuration, i.e. a second wing is placed downstream at a distance $\mathbf{r} = (0, 0, r_z)$ along the streamwise direction, as sketched in Fig.~\ref{fig:sketch-multiple}a.
We retain the initial perturbation only for the upstream wing to evidence how the downstream wing dynamics is affected by the impacting wake.
To accommodate a second downstream device, the domain is enlarged in the streamwise direction up to $z = 20$.

\begin{figure}
\centering
\includegraphics{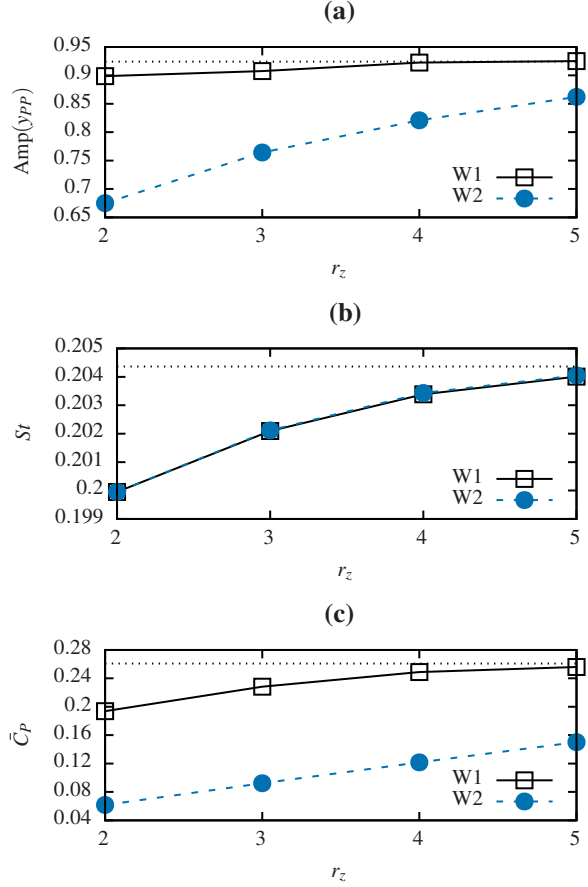}
\caption{
Flapping observables for in-line arrangement as a function of the distance between the two wings: (a) PP transverse oscillation amplitude, (b) Strouhal number and (c) power coefficient.
Solid line and empty squares: upstream wing (W1); dashed line and filled circles: downstream wing (W2); dotted line: values for the isolated, single device.
}
\label{fig:in-line_qoi-vs-r3}
\end{figure}

To start our analysis, we fix $K=3$ for both devices, since this was found to be the most efficient condition (i.e., maximising $\eta$) in the case of the single device, and perform simulations for different values of the relative distance $r_z$ (see Fig.~\ref{fig:sketch-multiple}a).
An insight of the resulting dynamics for the case $r_z=2$ is given by the top panels of Fig.~\ref{fig:in-line_TH-PS_tuned-K2}.
Looking at the time trace of the plunging motion (Fig.~\ref{fig:in-line_TH-PS_tuned-K2}a),
after a short transient (about 2 cycles)
a phase shift is established between the oscillations of the two devices.
In the new flapping state, the motion of the downstream wing gets synchronized to the wake released by the upstream one, the pitching motion being driven by low-pressure vortical regions, as it can be observed from Fig.~\ref{fig:in-line_vort}, showing instantaneous views of the vorticity field within one flapping cycle (see also Movie 1 provided with the Supplementary Material).
Furthermore, from the shape of limit-cycles reported in Fig.~\ref{fig:in-line_TH-PS_tuned-K2}b, we note that the state-space trajectory of the second device is contained within that of the upstream device.
 
Similar findings are obtained when varying the distance $r_z$, as shown in Fig.~\ref{fig:in-line_qoi-vs-r3}: the downstream device always oscillates with amplitude smaller than the upstream one (Fig.~\ref{fig:in-line_qoi-vs-r3}a), with a monotonic trend that seems to recover the single-device behaviour for large $r_z$, as expected. 
The flapping frequencies of the two devices, shown in Fig.~\ref{fig:in-line_qoi-vs-r3}b, are essentially locked and slightly smaller than that of the single case.
 $\mathit{St}$ decreases while decreasing the separation distance,
reflecting an alteration of the upstream wing dynamics, as well.
Finally, the power coefficient shows a similar trend (Fig.~\ref{fig:in-line_qoi-vs-r3}c), although the difference with respect to the single configuration is more pronounced: in fact, this quantity involves the product between the lift force and the PP velocity, both being weakened.

\subsubsection{Structural tuning of downstream device}

\begin{figure}
\centering
\includegraphics{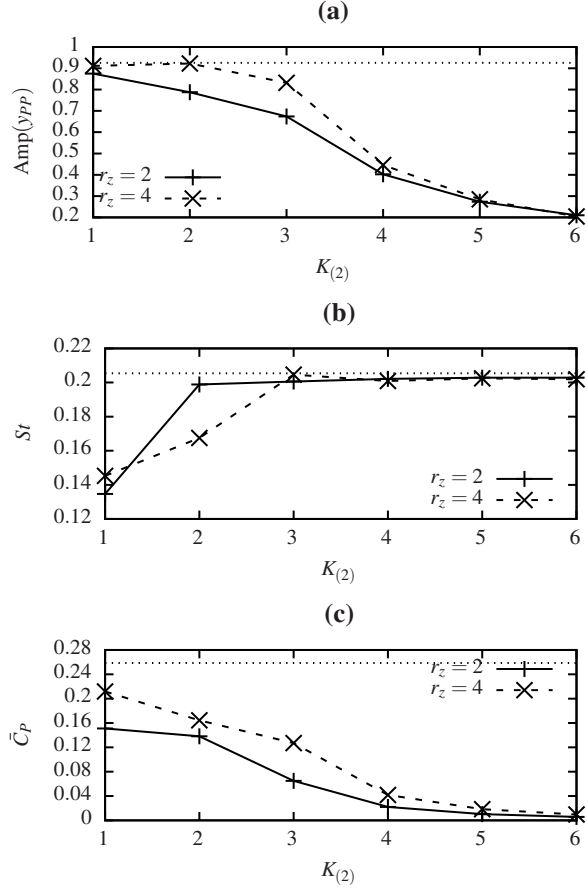}
\caption{ 
Flapping observables of downstream wing in in-line arrangement with tuning of its stiffness, as a function of this latter: (a) PP transverse oscillation amplitude, (b) Strouhal number and (c) power coefficient.
Solid line and  \textsf{+}'s: $r_z = 2$; dashed line and \textsf{x}'s: $r_z = 4$; dotted line: values for single-device configuration with $K=3$.
}
\label{fig:in-line_qoi-vs-K2}
\end{figure}

In order to improve the performance of the downstream device, we investigate the configuration in which the value of its stiffness $K_{(2)}$ is varied, while for the upstream device it is mantained fixed to $K_{(1)}=3$. Two streamwise distances are considered, i.e. $r_z = \{2,4\}$. Similarly to the previous case, the flapping observables are reported in Fig.~\ref{fig:in-line_qoi-vs-K2}.
The plunge motion increases its amplitude for decreasing $K_{(2)}$, attaining values close to the device in the single configuration (Fig.~\ref{fig:in-line_qoi-vs-K2}a). 
Conversely, the Strouhal number (Fig.~\ref{fig:in-line_qoi-vs-K2}b) decreases while softening the spring for $K_{(2)} < K_{(1)} = 3$. In this range, the resulting dynamics is a nonlinear combination of the self-excitation, taking place in uniform flow, and the wake-forcing mechanism previously discussed.

Bottom panels of Fig.~\ref{fig:in-line_TH-PS_tuned-K2} show the resulting flapping in time for a representative case, where one can notice the larger amplitude of both pitch and plunge compared to the corresponding case with $K_{2} = K_{1}$.
Furthermore, for this particular case ($r_z=2$, $K_{2} = 1$), one can see that the system shows asymmetric and multiperiodic oscillation of the pivot-point (Fig.~\ref{fig:in-line_TH-PS_tuned-K2}c,d).
Indeed, it was observed that when $K_{(2)} < K_{(1)}$ the spectral content is richer and the dominant frequency is altered compared to the single-wing case at $K = K_{(2)}$.

For $K_{(2)} > K_{(1)} = 3$, the flapping frequency remains locked to the upstream wing. The downstream-wing oscillation decreases while increasing $K_{(2)}$ but flapping now occurs also for $K_{(2)} > K_\mathrm{cr} \approx 4.7$, unlike what happens in the case of an isolated device. In this case, the only mechanism causing such motion is wake forcing.

A recovery in the value of the power coefficient is obtained by lowering $K_{2}$, as shown in Fig.~\ref{fig:in-line_qoi-vs-K2}c, which is beneficial from the EH perspective, although in none of the considered cases the same amount as for the single device was obtained.
Our numerical results can be compared with the experimental evidence recently presented in Ref.~\cite{kirschmeier2018wake}, where the concept of tuning the pitching stiffness of the downstream wing was proposed for pitch-and-plunge EH systems:
despite the different structural features (e.g., the presence of restoring moments both in plunging and pitching) and the significant difference in Reynolds numbers considered, 
a qualitative analogy can be drawn.

\begin{figure}
\centering
\includegraphics{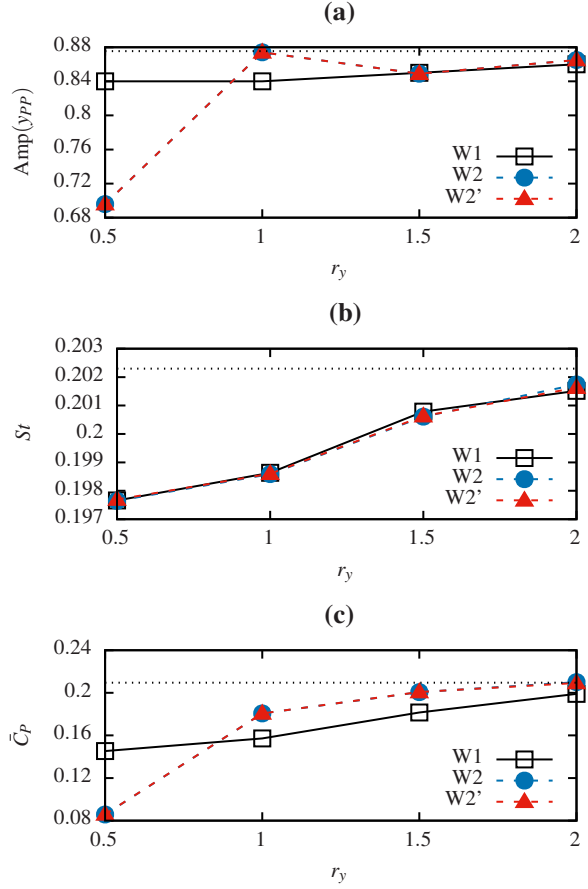}
\caption{ 
Flapping observables for staggered arrangement (Fig.~\ref{fig:sketch-multiple}b), as a function of the transverse distance between devices: (a) PP transverse oscillation amplitude, (b) Strouhal number and (c) power coefficient.
Black solid line and empty squares: W1; blue dashed line and filled circles: W2; red dashed line and filled triangles: W2'; black dotted line: values for single-device configuration with $K=K_{(1)}=3$.
}
\label{fig:staggered_qoi-vs-ry}
\end{figure}

\subsection{Staggered arrangement}
\label{sec:staggered}

In this case (Fig.~\ref{fig:sketch-multiple}b), a pair of devices (W2 and W2') is placed aft of the first one (W1) at a distance $\mathbf{r} = (0, \pm r_y, r_z)$.
We choose the closest streamwise distance considered for the in-line case, i.e. $r_z=2$, and investigate the behaviour for different values of the transverse distance $r_y = \{0.5,1,1.5,2\}$.

\begin{figure}
\centering
\includegraphics{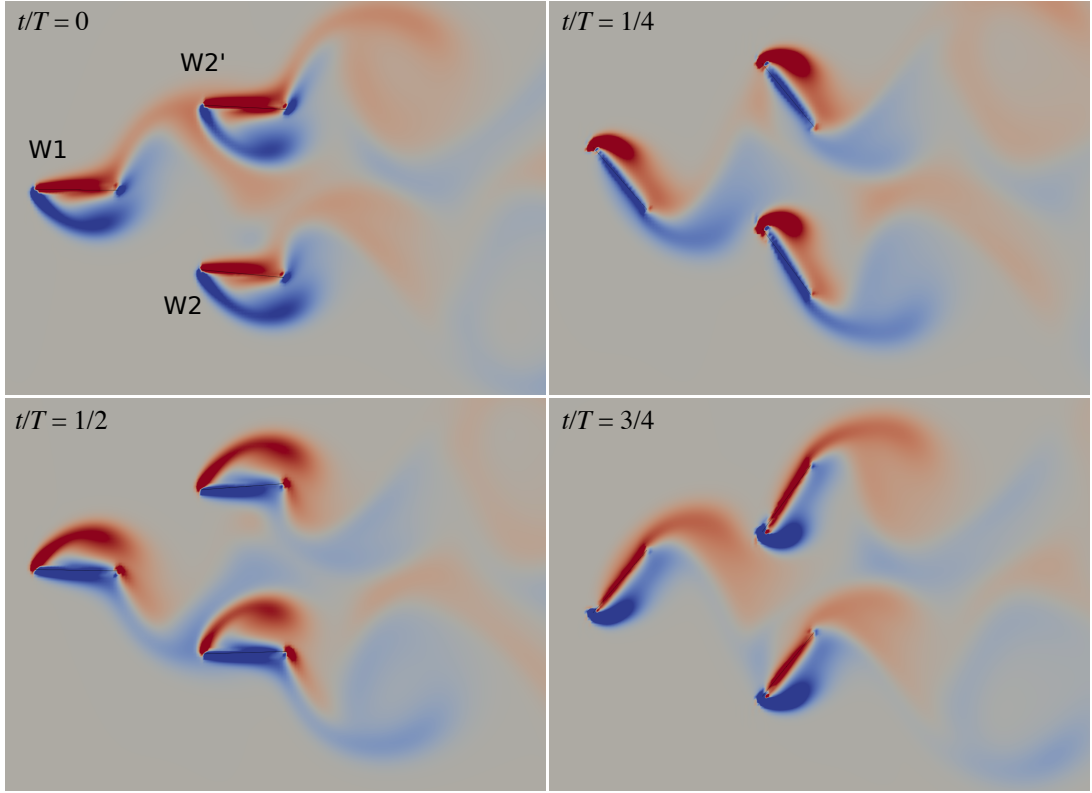}
\caption{
\added{Instantaneous views of plate position and vorticity field (negative values (i.e. counterclockwise) in blue, positive ones (i.e. clockwise) in red) within one flapping cycle for the staggered arrangement with $r_y = 1$ and $r_z = 2$.
}}
\label{fig:staggered_vort}
\end{figure}

Results of the analysis are presented in Fig.~\ref{fig:staggered_qoi-vs-ry}\added{.}
\added{
Concerning the plunging amplitude (Fig.~\ref{fig:staggered_qoi-vs-ry}a), an attenuation is found with respect to the single-device configuration for all devices, which is however limited for W1 to about $5\%$, while for W2 and W2' we have a relative peak at $r_y = 1$ (approaching the value in single configuration) and a sharp decrease for $r_y = 0.5$ (about $20\%$).
In the latter, the downstream wings are found to lie entirely within the wake released by W1, the oscillation being reduced in a way similar to what reported for the in-line arrangement, featuring the same synchronization mechanism.
For $r_y=1$, the wake is impacting on downstream devices only during a portion of the flapping cycle, yielding a different dynamics: the flapping motions are collectively in-phase, as shown by Fig.~\ref{fig:staggered_vort}, and the occurring interference is presumably responsible for the relative peak in amplitude. 
Increasing the distance to $r_y=1.5$ and 2, the same in-phase motion is still observed but W2 and W2' are now located outside the wake region and consequently the weakly constructive interference is not present anymore.
}

 Note that for W2 and W2'  the flapping motion is asymmetric with respect to the streamwise direction but substantially specular with respect to each other.
The Strouhal number shows almost negligible variations (less than $5\%$) compared to the single-device configuration, decreasing for smaller $r_y$ (Fig.~\ref{fig:staggered_qoi-vs-ry}b).
Values of the power coefficient are always found to be lower than that of the isolated device.
We note however that, for $r_y > 1$, the power coefficient is found to be higher for downstream devices (Fig.~\ref{fig:staggered_qoi-vs-ry}c).
\added{}

\subsection{Side-by-side arrangement}
\label{sec:side-by-side}

\begin{figure}
\centering
\includegraphics{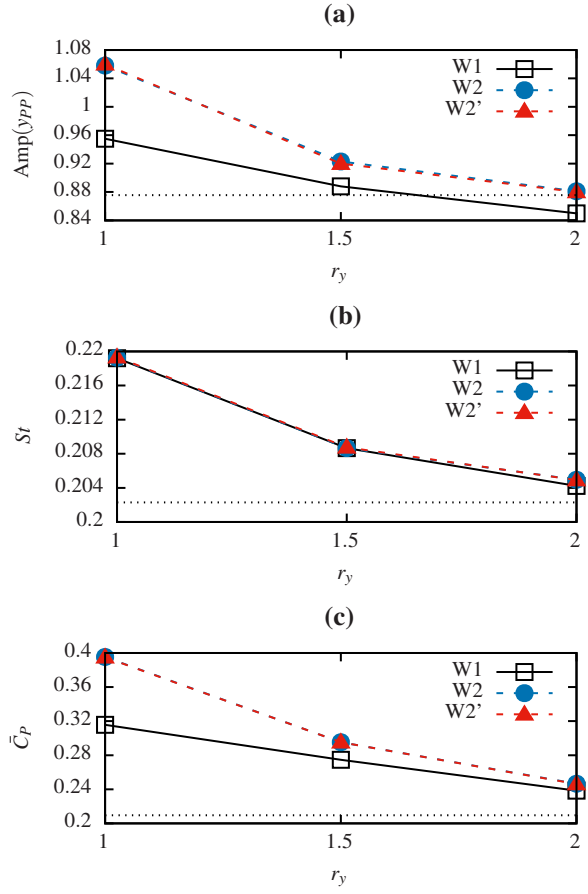}
\caption{ 
Flapping observables for side-by-side arrangement (Fig.~\ref{fig:sketch-multiple}c), as a function of the transverse distance between devices: (a) PP transverse oscillation amplitude, (b) Strouhal number and (c) power coefficient.
Black solid line and empty squares: W1; blue dashed line and filled circles: W2; red dashed line and filled triangles: W2'; black dotted line: single-device configuration.
}
\label{fig:side-by-side_qoi-vs-ry}
\end{figure}

\begin{figure}
\centering
\includegraphics{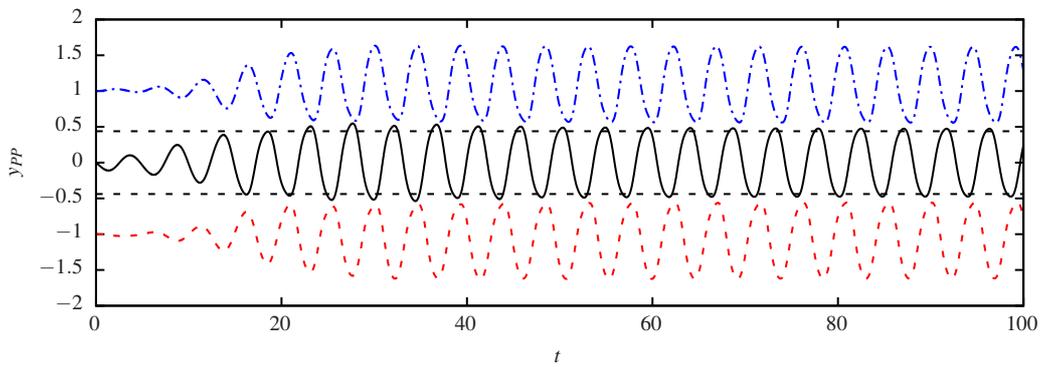}
\caption{Time history of transverse PP oscillation for devices in side-by-side arrangement with $r_y=1$. Black solid line:  W1; red dashed line:  W2; blue dot-dashed line: W2'. Black dashed lines report the maximum and minimum of PP oscillation in the single configuration.
}
\label{fig:side-by-side_TH_yPP}
\end{figure}

We now focus on the side-by-side arrangement, where we place a pair of devices (W2 and W2') at a transverse distance $\mathbf{r} = (0, \pm r_y, 0)$ with respect to the central one (W1), as sketched in Fig.~\ref{fig:sketch-multiple}c.
As for the other two arrangements, only the first wing is initially perturbed.

Flapping observables from numerical simulations considering three different values of the mutual distance, i.e. $r_y = \{1,1.5,2\}$, are reported in Fig.~\ref{fig:side-by-side_qoi-vs-ry}.
Looking at the plunging amplitude (Fig.~\ref{fig:side-by-side_qoi-vs-ry}a), for sufficiently small separations (i.e. $r_y \leq 1.5$) the resulting values are larger than that obtained for the single device, although the relative increment is only up to about $8\%$ for W1 and $20\%$ for W2 and W2'. 
The flapping frequency is nearly identical for the three devices and slightly increases while decreasing $r_y$, up to about $5\%$ when $r_y=1$ (Fig.~\ref{fig:side-by-side_qoi-vs-ry}b).
Despite the relatively small variation of these two quantities, a significant increase of the power coefficient occurs for all flapping plates (Fig.~\ref{fig:side-by-side_qoi-vs-ry}c):
with respect to the single device, for the central wing the increase is almost $50\%$, while for the side wings this reaches nearly $90\%$.

Similarly to what it was observed for the staggered arrangement (Sec.~\ref{sec:staggered}), the central wing undergoes symmetrical motion while the side wings exhibit slightly asymmetrical flapping,
as it can be observed from the time traces of the transverse PP displacement reported in Fig.~\ref{fig:side-by-side_TH_yPP} (see also Movie 2 in the Supplementary Material).
Moreover, the steady-state oscillations of the side wings are approximately in-phase with respect to each other and in counter-phase with respect to the central wing.
From the qualitative viewpoint, we can observe how the resulting scenario resembles the counter-phase flapping regime observed for flexible bodies placed at moderate distances~\cite{zhang2000flexible,favier2015,huertas-cerdeira2018}.

\begin{figure}
\centering
\includegraphics{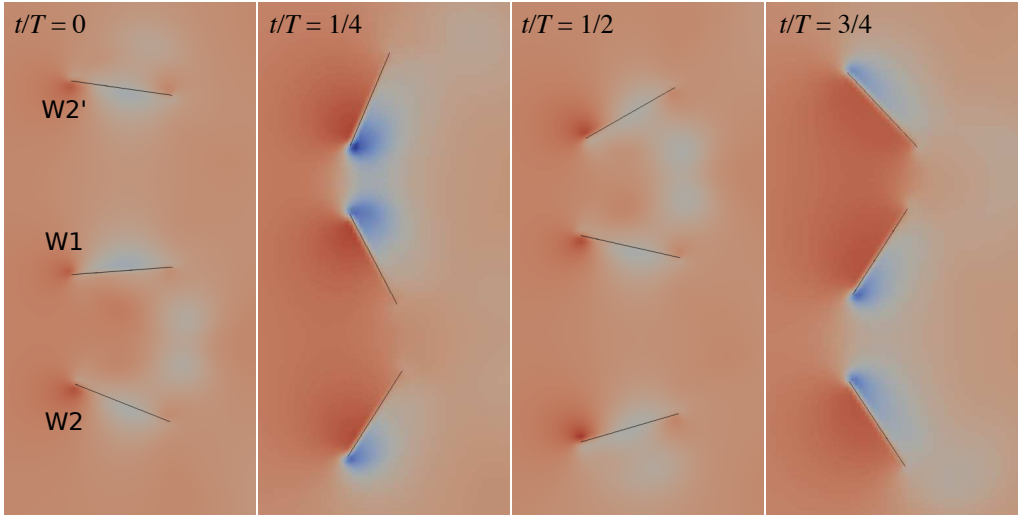}
\caption{Instantaneous views of plate position and pressure field (negative values in blue, positive ones in red) within one flapping cycle for the side-by-side arrangement with $r_y = 1.5$.
}
\label{fig:side-by-side_p}
\end{figure}

The described dynamics is supported by Fig.~\ref{fig:side-by-side_p}, where the position of the three wings along with the pressure field at different instants within one flapping cycle is shown. From these snapshots, we argue that when two adjacent wings get closer (approximately when the pitching angle is at its maximum/minimum), the flow velocity has to increase due to the narrower effective cross section; hence, the pressure minimum gets amplified compared to the single case, this in turns increasing the amplitude of the lift force and, consequently, the wing oscillation. 

\begin{figure}
\centering
\includegraphics{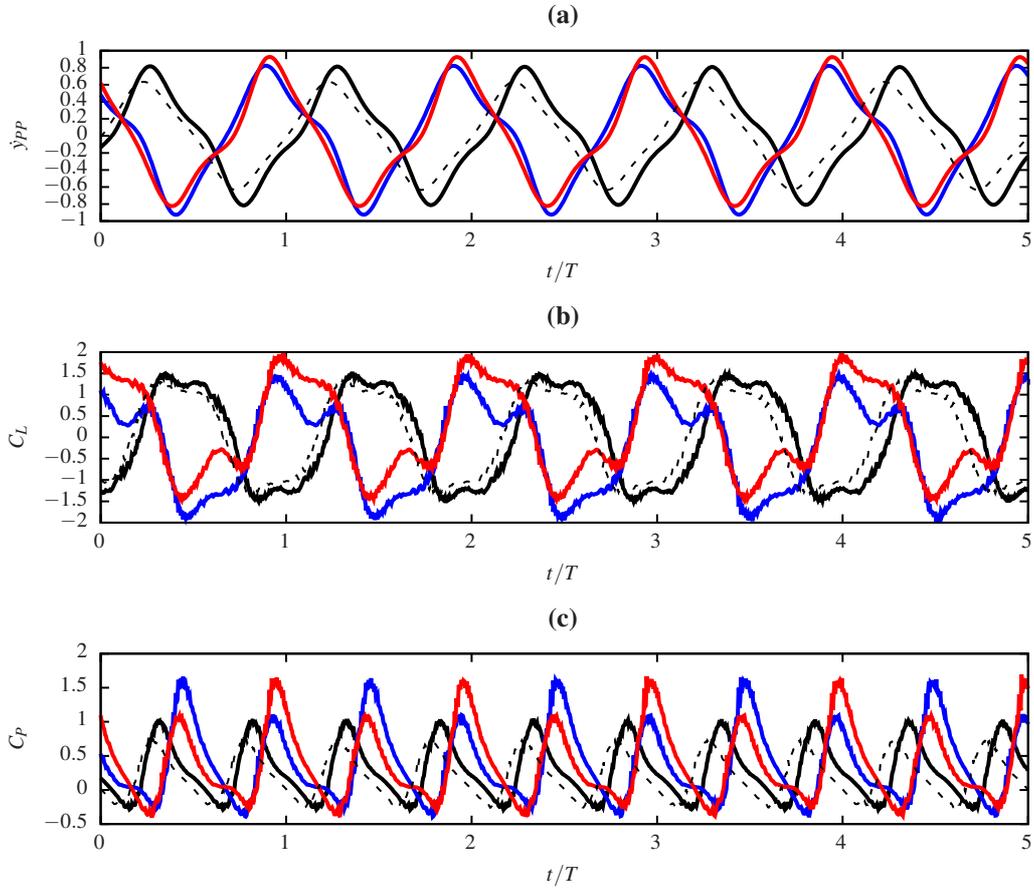}
\caption{Time histories of (a) PP transverse velocity, (b) lift coefficient and (c) instantaneous plunge power coefficient, \added{comparing the central (W1, black), bottom (W2, blue) and top (W2', red) wings in side-by-side arrangement with $r_y=1$ (solid line) and the same device in single configuration (dashed line). Time is normalized by the corresponding period of flapping motion.
}}
\label{fig:side-by-side_TH_cfr}
\end{figure}

In Fig.~\ref{fig:side-by-side_TH_yPP}, the PP oscillation amplitude of W1 is compared to that of the isolated case. As mentioned before, the variation of this quantity does not provide a direct indication of the gain in performance that is achieved employing this configuration.
To perform a more detailed comparison, \added{let us focus on the quantities that are directly associated with the plunge power, whose time evolutions are reported in Fig.~\ref{fig:side-by-side_TH_cfr}.
The first one is the PP transverse velocity (Fig.~\ref{fig:side-by-side_TH_cfr}a), for which the observed difference, in terms of amplitude, is about $20\%$ for all wings.
The second is the lift coefficient
 (defined as $C_L = 2 \,F_\mathrm{aero}^y / (\rho_\mathrm{f} c s U^2)$), shown in Fig.~\ref{fig:side-by-side_TH_cfr}b, for which the variation is approximately $20\%$ for W1 and $40\%$ for the side wings.
 For the latter, the increment is even more pronounced since the flapping motion is asymmetric, so that higher positive (or negative) peaks of $C_L$ are found for the top (or bottom) wing. 
By multiplying these two quantities, we obtain the instantaneous power coefficient reported in Fig.~\ref{fig:side-by-side_TH_cfr}c, where one can see how the positive peaks in the side-by-side configuration are appreciably higher than in the single case, while the negative peaks remain essentially at the same values. Consequently, the average power coefficient undergoes the significant increase already presented in Fig.~\ref{fig:side-by-side_qoi-vs-ry}c.
}

\begin{figure}
\centering
\includegraphics{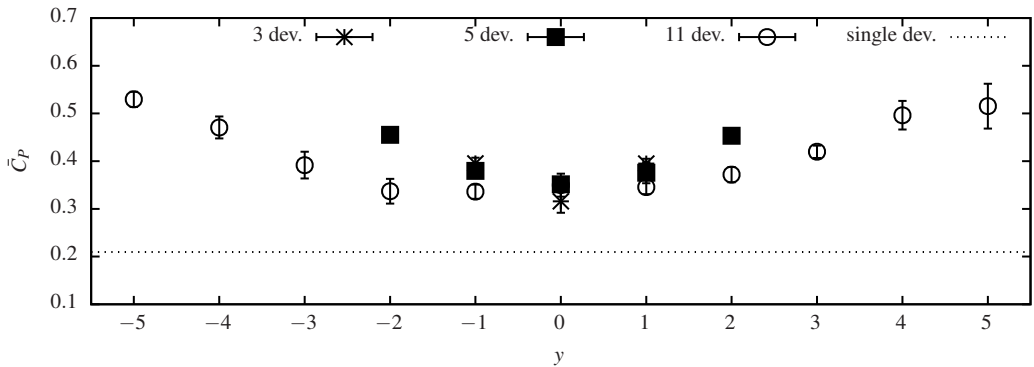}
\caption{Power coefficient distribution for side-by-side arrangements with different number of devices, placed at mutual transverse distance $r_y=1$.
Error bars indicate the variation in the performed cumulative average due to non-regular flapping.
}
\label{fig:side-by-side_comp-Ndev}
\end{figure}

In light of these results, we move further by considering arrays of more than three objects, fixing the mutual distance between adjacent devices to $r_y = 1$, for which we found the highest increase in performance.
The outcome of this analysis is presented by Fig.~\ref{fig:side-by-side_comp-Ndev} collecting the average power coefficient for each device: when placing two additional plates at $y = \pm 2$ (i.e., considering an array of $N_\mathrm{d}=5$ devices), these achieve a further enhanced performance compared to the case where $N_\mathrm{d}=3$.
The same trend holds when increasing the number of devices to $N_\mathrm{d}=11$.
Increasing $N_\mathrm{d}$, the motion of flapping objects may be found to be less regular and periodic, as e.g. for the side wing placed at $y=5$. Hence, we now evaluate the average in a statistical sense, assessing its convergence in time (this corresponding to error bars in  Fig.~\ref{fig:side-by-side_comp-Ndev}).

Emphasizing the comparison with the total power expected considering $N_\mathrm{d}$ isolated devices, we can write the overall power coefficient associated with the whole array $\bar{C}_P^\mathrm{tot}$ as:
\begin{equation}
  \bar{C}_P^\mathrm{tot}  = \sum_{i=1}^{N_\mathrm{d}} \bar{C}_P =   {N_\mathrm{d}} \bar{C}_P^\mathrm{S} + I,
 \label{eq:CP_array}
\end{equation}
where $\bar{C}_P^\mathrm{S}$ is the power coefficient of the single device in isolated configuration and the interference quantity $I=I(N_\mathrm{d})$ has been introduced.
The latter quantifies the additional power due to the cooperative effect.
Similarly, the relative increment can be expressed by introducing the interference factor
 \begin{equation}
 IF =  \frac{I}{N_\mathrm{d} \bar{C}_P^\mathrm{S}}.
 \label{eq:interferenceFactor}
\end{equation}

Table~\ref{tab:gain_array} collects the values obtained in our computational study, from which $\mathit{IF}$ seems to converge with the number of devices approximately to unity, i.e. the amount of available power is almost $100\%$ increased compared to the sum of ${N_\mathrm{d}}$ isolated devices.
Moreover, this evidence can be representative when extrapolating our results to the situation of a network made of a vast number of devices.
In this situation the side effects, although beneficial in terms of performance, are expected to be negligible.
Nevertheless, each device can be thought to behave as those in the central part of the array. 
We can therefore conclude that employing a side-by-side arrangement of many devices appears to be substantially beneficial in terms of EH potential.

\begin{table}[tp]
\caption{Performance of side-by-side arrays with different number of devices $N_\mathrm{d}$; $\bar{C}_P^\mathrm{tot}$ is the total power coefficient of the array (Eq.~\eqref{eq:CP_array}) and $IF$ is the interference factor as defined by Eq.~\eqref{eq:interferenceFactor}.}
\begin{center}
\begin{tabular}{l l l}
\toprule
$N_\mathrm{d}$ & $\bar{C}_P^\mathrm{tot}$ & $IF$ \\
\midrule
3 & 1.10 & 0.76 \\
5 & 2.01 & 0.92 \\
11 & 4.56 & 0.98 \\
\bottomrule
\end{tabular}
\end{center}
\label{tab:gain_array}
\end{table}

\subsubsection{Wind-tunnel experiments}

\begin{figure}
\centering
\includegraphics[width=0.5\textwidth]{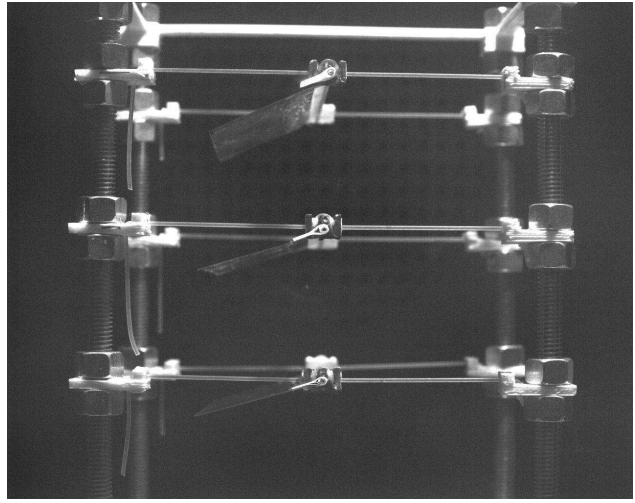}
\caption{
Experimental realization of side-by-side array made by three devices exploiting elastically-bounded plates (similar to those of Refs.~\cite{olivieri2017fluttering,olivieri2017aeroelastic}).
}
\label{fig:exp_array}
\end{figure}

In order to corroborate the numerical findings, our study is complemented by an experimental analysis of the side-by-side arrangement made of an array of three devices, shown in Fig.~\ref{fig:exp_array}. 
Experiments were performed using the wind-tunnel facility at Physics Department of the University of Genoa, whose characteristic parameters, along with the procedure used to acquire the motion of flapping wings in time, have been described in previous works~\cite{boragno2012elastically,olivieri2017fluttering,boccalero2017power}.

Each of the three components is similar to the devices already presented in Refs.~\cite{olivieri2017fluttering,olivieri2017aeroelastic} and is briefly described as follows. 
A $0.1\mathrm{mm}$-thick foil made of polyvinil acetate, with planform dimensions  $c=30\mathrm{mm}$ and $s=65\mathrm{mm}$, is glued to a brass rod that is connected (allowing free rotation) to four elastomeric elements
made of siliconic rubber, aligned with the streamwise direction and pre-stretched so that, 
following Ref.~\cite{olivieri2017fluttering}, the value of an effective stiffness can be estimated to be $\mathcal{K}^\mathrm{eff}_{(y)} \approx 13 \mathrm{N/m}$.
The mass of the moving body is $m= 0.94\mathrm{g}$ and the center of mass is approximately placed $0.2c$ aft of the leading edge.

\begin{figure}
\centering
\includegraphics{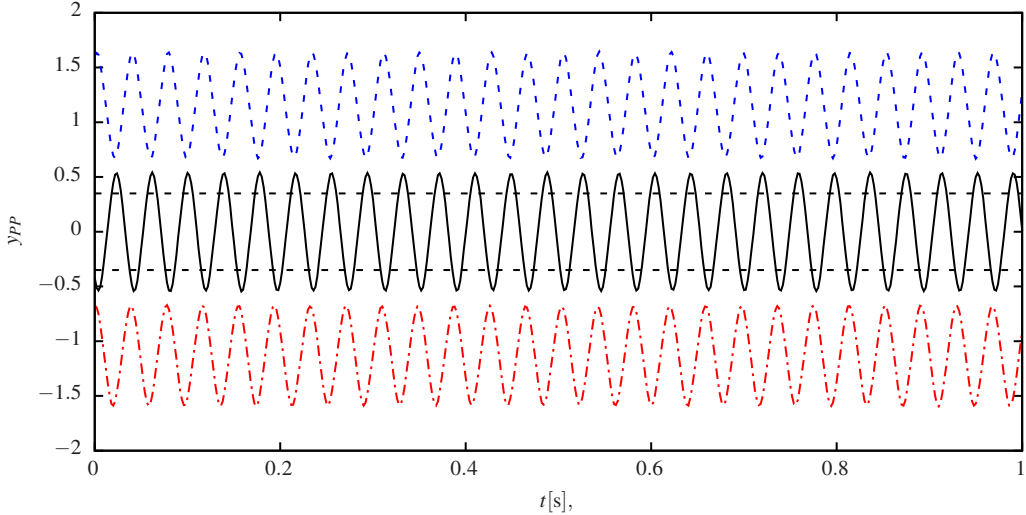}
\caption{Time history of transverse PP oscillation for the experimental array of three devices in side-by-side arrangement (Fig.~\ref{fig:exp_array}). Black solid line:  W1; red dot-dashed line:  W2; blue dashed line: W2'. Black dashed lines indicate maximum and minimum displacement of W1 in the single configuration.
}
\label{fig:side-by-side_TH_exp}
\end{figure}

The three devices are collocated within a fixed frame and placed approximately at the center of the test chamber. The mutual distance between devices corresponds to $r_y \approx 1.1$.
A freestream air velocity $U=4.5\mathrm{m/s}$ is chosen, at which regular LCO was found for devices tested in the single configuration. 

Assuming air at standard conditions, the chord-based Reynolds number can be estimated as $\mathit{Re} \approx 9 \times 10^3$, while
from the other quantities we can derive the remaining equivalent nondimensional parameters: 
$\AR \approx 2$, $\rho_\mathrm{w} \approx 17$ and $K \approx 21$.
Except for the aspect ratio, these values are clearly different from those considered in the numerical investigation. The analysis is thus intended as complementary, in order to assess the robustness of the outlined mechanism when moving into the operational range of the real EH application.

The experiment has been conducted as follows. First, we performed measurements of each device taken individually, i.e. removing the other two from the array and retaining the device under consideration in the same position as in the multiple configuration. 
Then, we tested the multiple configuration, where all three devices are present.

\begin{figure}
\centering
\includegraphics[width=0.8\textwidth]{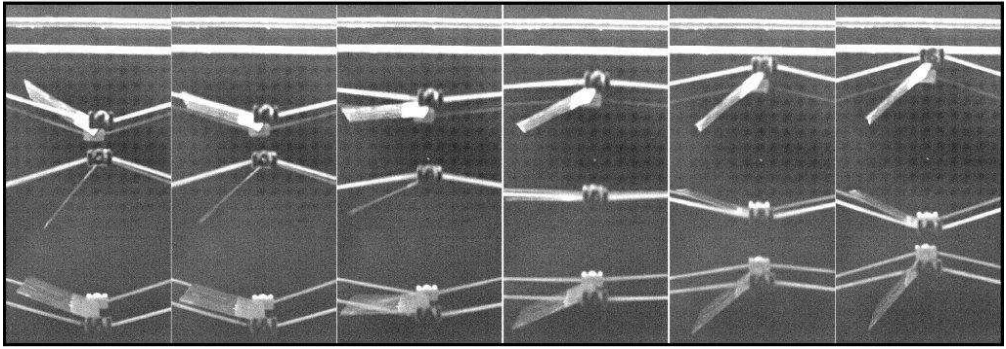}
\caption{
Side views of the experimental side-by-side arrangement of three devices during collective flapping motion (the wind is coming from the right and the time interval between frames is $4\mathrm{ms}$).
}
\label{fig:exp_snapshots}
\end{figure}

\begin{table}[tp]
\caption{Flapping observables for the experimental array of three devices in side-by-side arrangement. The suffix denotes values for the single (S) or the array (A) configuration.}
\begin{center}
\begin{tabular}{l l l l l l l}
\toprule
Device & $\mathrm{Amp}(y_\mathit{PP})^\mathrm{S}$ & $\mathrm{Amp}(y_\mathit{PP})^\mathrm{A}$ & $\mathit{St}^\mathrm{S}$ & $\mathit{St}^\mathrm{A}$ & $\bar{C}_P^\mathrm{S}$ & $\bar{C}_P^\mathrm{A}$\\
\midrule
W2' & 0.69 & 0.97 & 0.16 & 0.17 & 0.13 & 0.30 \\
W1 & 0.73 & 1.07 & 0.16 & 0.17 & 0.16 & 0.43 \\
W2 & 0.83 & 0.9 & 0.17 & 0.17 & 0.23 & 0.29 \\
\bottomrule
\end{tabular}
\end{center}
\label{tab:exp_single}
\end{table}

The time traces of the acquired PP oscillation for each device are reported in
Fig.~\ref{fig:side-by-side_TH_exp}, where it can be observed that the behaviour is  analogous to that of numerical simulations:
the amplitude of the oscillation increases and
motions of adjacent wings are essentially in counterphase, as it is also shown by the side views of Fig.~\ref{fig:exp_snapshots} (see also Movie 3 in the Supplementary Material).

To quantify the effect due to the mutual interaction, Table~\ref{tab:exp_single} reports values of the same observables analysed numerically, i.e. amplitude of PP oscillation, Strouhal number and power coefficient, both for the single and multiple configurations.
Focusing on data of the single configuration, we observe that a certain difference exists between the three devices, which can be ascribed mainly to constructive details.
Nevertheless, when considering the same quantities in the multiple configuration, 
an increase with respect to the single case is found for all devices.
In particular, the PP amplitude is maximised for the central wing (W1), with an increase around $40\%$ with respect to the individual configuration.
Variation in frequency is smaller, i.e. between $2$ and $10\%$ with the same resulting synchronisation between devices that is observed numerically.
From the acquired movies it is also possible to evaluate the aerodynamic forces and thus the power associated to the plunge and pitch motions.
In Table~\ref{tab:exp_single}, we report the cycle-averaged plunge power coefficient, as defined in Eq.~\eqref{eq:CP}.
The increment when the three wings are coupled is evident, confirming qualitatively the results obtained by the simulations.

\added{From the quantitative viewpoint, some differences can be noted between the experimental and the numerical results (i.e., comparing Table~\ref{tab:exp_single} with Fig.~\ref{fig:single} for the single device and Fig.~\ref{fig:side-by-side_qoi-vs-ry} for the side-by-side arrangement).
We underline, however, that the experimental configuration and the idealized model differ in several features that may explain the observed discrepancies: for example, for the former we have: (i) higher $\mathit{Re}$ number, (ii) nonhomogeneous mass distribution and (iii) nonlinear elastomeric elements in place of Hooke springs.
These features appear to influence the device dynamics and on-going investigations are aimed at characterizing more exhaustively their role. In particular, we plan to perform simulations for a configuration where such details are included, hence further more representative of the real EH device.}

\section{Conclusion}
\label{sec:conclusions}

This work has investigated the dynamics of an aeroelastic system suitable for EH purposes, based on fluttering oscillations of elastically-bounded plates in a laminar flow -- focusing, in particular, on the interaction between multiple devices arranged into three basic configurations (i.e.: in-line, staggered and side-by-side).
As a first step, we have considered an effective physical model where homogeneous plates are anchored by linear springs and immersed in uniform flow. By employing a suitable numerical procedure based on a finite-difference Navier-Stokes solver coupled with a moving-least-squares immersed boundary method, we have performed numerical simulations for the single and multiple wing configurations.
Focusing on the main observables of flapping motion such as the amplitude and frequency of oscillation, as well as performance-related quantities (i.e. power coefficient and Betz efficiency), we investigated how the dynamical behaviour is affected by the governing parameters (e.g., the nondimensional stiffness) and the mutual distance at which devices are placed in the multiple configuration.

For both the in-line and staggered arrangements, the performance of interacting devices is found to be worse than in the single case, although a recovery in performance can be achieved by tuning the elasticity of downstream devices.
When considering the side-by-side configuration, in contrast, the interaction turns out to be beneficial and relevant increases of all quantities of interest are found.
Further enhancements are obtained when increasing the number of devices in the array. A constructive interference is found, causing the total power coefficient of the network to increase up to $100\%$ with respect to the expected amount by the same number of isolated devices.

The occurrence of this cooperative effect has been verified experimentally in a case representative of the real EH application, considering an array of prototypal devices in a higher Reynolds flow, revealing good agreement with numerical findings despite the different configurations.
The present results therefore suggest the development of dense arrays of flutter-based EH devices as a strategy that could enable higher performance for this kind of novel technology.

\section*{\added{Acknowledgements}}
\added{The authors acknowledge CINECA and INFN for the availability of high performance computing resources and support.}

\appendix

\section{Validation of numerical method}
\label{sec:validation}

The numerical procedure used in the present study has been extensively validated for several FSI problems involving both rigid and deformable bodies~\cite{detullio2016,spandan2017}.
In the current investigation, however, we assess the dependency of numerical results on the spatial and temporal resolution.
To this end, the chord-based Reynolds number is set to $\mathit{Re}=1000$, i.e. the highest value tested in the whole study, and we choose the baseline values for the other control parameters, i.e.  $\AR=2$, $\rho_\mathrm{w} = 2$ and $K=3$, this configuration corresponding to a regular flapping state of finite amplitude. A finite initial perturbation on the angular velocity is given in order to shorten the transient, although the achievement of the same steady limit-cycle was verified in case of unperturbed initial condition.

The wing is initially placed with its geometrical center at the origin $(0, 0, 0)$.
The domain dimensions are set at $\pm5$ both in the $x$ and $y$ direction, and at $-5$ and $10$ in the $z$ direction.
A Cartesian grid is used, with uniform resolution in the spanwise direction and stretched in the other two directions. However, the grid spacing $h$ is made locally uniform in a region close to the wing, i.e. for $-1 < z < 2$ and $-1 < y < 1$. 
In order to quantify convergence with respect to the spatial resolution, eleven different grids were considered, characterized as shown in Table~\ref{tab:gc_grids}. 
While the same topology is retained, the grids differ by an overall refinement factor. 
The wing is assumed to have zero thickness and is thus discretized in a surface mesh made of triangular elements whose characteristic length is proportionally adjusted to be essentially equal to the minimum Eulerian spacing $h$.

\begin{table}[tp]
\caption{Grid settings used for convergence study. $N_x$, $N_y$ and $N_z$ denote the number of nodes in the $x$, $y$ and $z$ direction, while $H$ and $h$ indicate the minimum and maximum resolution, respectively.}
\begin{center}
\begin{tabular}{l l l l l l}
\toprule
Grid & $N_x$ & $N_y$ & $N_z$ & $H$ & $h$ \\
\midrule
A & 251 & 84 & 127 & $0.4$ & $0.04$\\
B & 271 & 97 & 147 & $0.35$ & $0.035$\\
C & 334 & 113 & 170 & $0.3$ & $0.03$\\
D & 401 & 136 & 205 & $0.25$ & $0.025$\\
E & 501 & 170 & 257 & $0.2$ & $0.02$\\
F & 541 & 189 & 286 & $0.18$ & $0.018$\\
G & 601 & 213 & 322 & $0.16$ & $0.016$\\
H & 641 & 235 & 353 & $0.146$ & $0.0146$\\
I & 751 & 258 & 388 & $0.133$ & $0.0133$\\
J & 865 & 298 & 448 & $0.115$ & $0.015$\\
K & 1001 & 344 & 516 & $0.1$ & $0.01$\\
\bottomrule
\end{tabular}
\end{center}
\label{tab:gc_grids}
\end{table}

The approximate solution can be expressed as a function of grid spacing as follows:
\begin{equation}
  f(h) \approx f^* + C h^p
 \label{eq:true_error}
\end{equation}
where $h$ is the grid spacing, $f^*$ is the exact solution (i.e., the solution that one would have for $h \to 0$), $C$ is a constant and $p$ is the (actual) order of convergence.
Instead of  fitting the obtained data using this expression,
we find more convenient to introduce the relative error with respect to the finest grid (for which $h=h_\mathrm{min}$), defined as
\begin{equation}
  \epsilon(h) = \frac{f(h) - f(h_\mathrm{min})}{f(h_\mathrm{min})},
 \label{eq:rel_error}
\end{equation}
since by combining Eqs.~(\ref{eq:true_error}) and~(\ref{eq:rel_error}), the following relation can be found:
\begin{equation}
  \epsilon(h) = C (h^p -  h_\mathrm{min}^p),
 \label{eq:rel_error_fit}
\end{equation}
where only $C$ and $p$ appear.
The fitting is therefore applied to this latter expression, from which we later extrapolate the exact value $f^*$.

\begin{figure}
\centering
\includegraphics{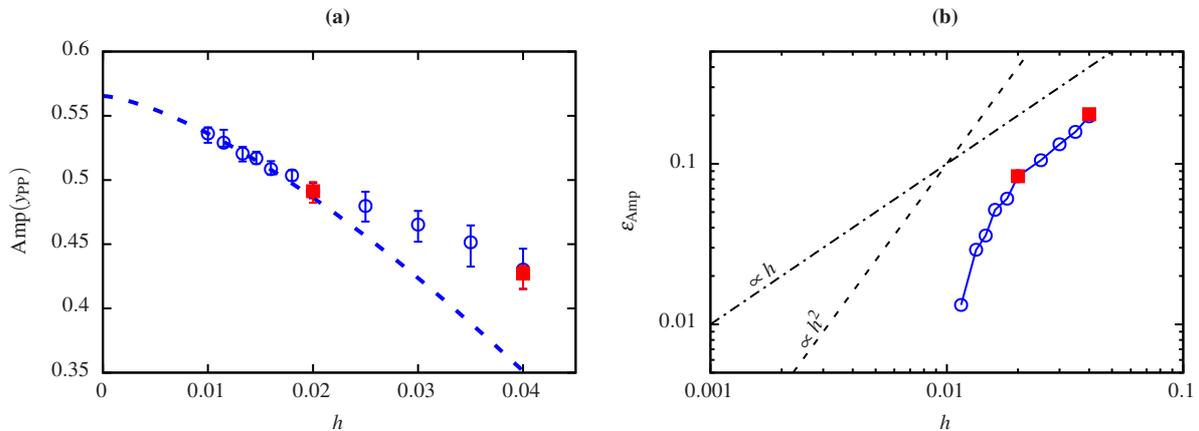}
\caption{
 Grid refinement analysis:
(a) amplitude of the pivot point transverse oscillation and (b) corresponding relative error with respect to the solution from the finest grid K (see Table~\ref{tab:gc_grids}), as a function of spatial resolution.
Blue circles correspond to cases with nonuniform grid spacing (listed in Table~\ref{tab:gc_grids}), from which the fitting curve indicated by the blue dashed line is derived, while red squares refer to cases with uniform grid spacing.}
\label{fig:gc_yPP}
\end{figure}

Grid convergence is assessed by considering the amplitude of the transverse oscillation of the pivot point \sloppy{$\mathrm{Amp}(y_\mathit{PP})$}.
Each case is computed up to measure at least 5 cycles of steady LCO.
Fig.~\ref{fig:gc_yPP}a shows the corresponding mean values along with the minimum and maximum ones.
From the plot one can notice that the convergence trend has a change at about $h = 0.02$, the actual order improving while increasing the resolution. 
By applying Eq.~\eqref{eq:rel_error_fit}, the curve depicted in Fig.~\ref{fig:gc_yPP}a is found, where the coefficients appearing in the expression are $C \approx -40.7$ and $p\approx 1.44$, yielding the extrapolated value of the exact solution $f^* \approx 0.56$.
In the plot we also report data from two cases employing uniform grids with constant spacing $h = \{0.02, 0.04 \}$, showing that the discrepancy with equivalent stretched grids (with same $h$) looks contained and decreases while increasing the resolution.
Moreover, we can also refer directly to the relative error with respect to the finest grid considered, as defined by Eq.~\eqref{eq:rel_error}.
This quantity is shown in Fig.~\ref{fig:gc_yPP}b where we can notice overall second-order accuracy for finer grids and first-order accuracy for coarser ones (in agreement with previous studies~\cite{detullio2016}).

In the presented parametric studies, two different grid settings are used.
For the single (Sec.~\ref{sec:single}) and in-line (Sec.~\ref{sec:in-line}) configurations, grid E is chosen. 
For the staggered (Sec.~\ref{sec:staggered}) and side-by-side (Sec.~\ref{sec:side-by-side}) configurations, we use instead a uniform grid with $h=0.04$. The latter choice is motivated by the fact that a wider region of the domain has to be refined in this case. 
One can note that the oscillation gets underestimated when using coarser grids, thus the numerical result is generally conservative.
A similar trend can be found for the pitching amplitude,
while the flapping frequency shows a convergence that is far more rapid with variations less than $2\%$ within the considered range of spatial resolution.

\begin{figure}
\centering
\includegraphics{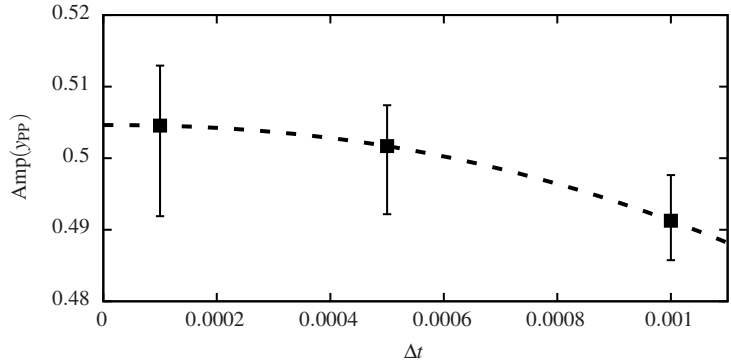}
\caption{
Amplitude of the PP transverse oscillation as a function of the numerical timestep, for the case of nonuniform grid E.
}
\label{fig:td_yPP}
\end{figure}

Fixing the choice of the grid, we are able to investigate the dependence of the solution with respect to the simulation timestep. Fig.~\ref{fig:td_yPP} shows the convergence when decreasing the timestep, with an estimated error of about 3\% with respect to the extrapolated value for vanishing $\Delta t$.
Last, the sensitivity to the domain size was checked, with negligible differences observed in the resulting flapping motion when doubling the size of the bounding box in all directions.

\begin{figure}
\centering
\includegraphics{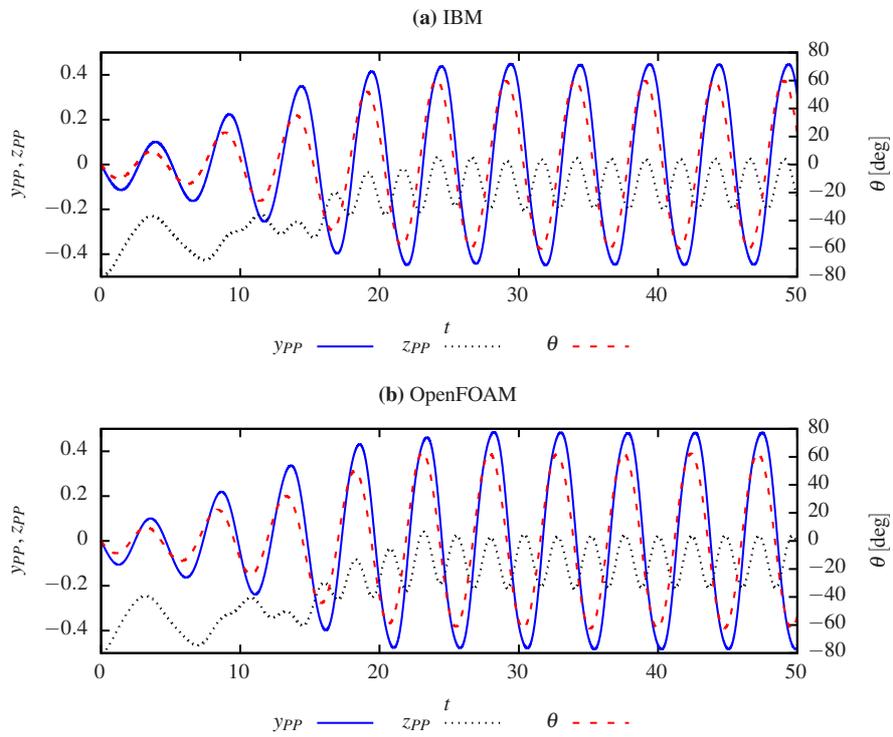}
\caption{
Comparison of the resulting plate motion between (a) the presently used immersed boundary method and (b) the body-conforming, overset mesh procedure, for the case $\mathit{Re}=100$, $\AR=2$, $\rho_\mathrm{w}=2$ and $K=3$ and the following observables: PP transverse translation (blue solid line), PP streamwise translation (black dotted line) and pitching angle (red dashed line).
}
\label{fig:cfr_OF}
\end{figure}

We conclude by providing a comparison of our results with those obtained with an essentially alternative approach, i.e. using a body-conforming mesh. The computation is performed using the open source library \textit{OpenFOAM}~\cite{of2018}, which is based on the finite volume method and offers dynamic mesh features, such as the overset treatment that is employed in our benchmark.
For this test, we set the case so that the spatial and temporal resolutions are comparable to those of grid A.
In Fig.~\ref{fig:cfr_OF}, we report the time histories of the pivot point motion and pitching angle obtained with the two approaches: all quantities look in good agreement, with small differences in the amplitude and period of the oscillation.

The simulations were performed on the same workstation using 8 processors. 
The comparison between the computational times of the two codes indicates a wall clock time of $\simeq 0.25\mathrm{s}$ per time step
for the IBM code and $\simeq 5\mathrm{s}$ for OpenFOAM, thus yielding a factor
20. Although the specific figures might depend on the particular
computer architecture and the details of the problem, it is clear that
the IBM code performs more efficiently for this class of problems.

It is worth mentioning that the differences might become even more evident when more than one flapping element is considered in the problem. In fact, while for IB methods the CPU time would grow only because of the larger number of immersed surface elements, for a moving grid method there would be more grid patches in relative motion and the computational load would increase more rapidly than the surface element counting.

\clearpage

%
  \bibliographystyle{elsarticle-num} 
  \bibliography{references}

\begin{thebibliography}{10}
\expandafter\ifx\csname url\endcsname\relax
  \def\url#1{\texttt{#1}}\fi
\expandafter\ifx\csname urlprefix\endcsname\relax\def\urlprefix{URL }\fi
\expandafter\ifx\csname href\endcsname\relax
  \def\href#1#2{#2} \def\path#1{#1}\fi

\bibitem{li2016review}
D.~Li, Y.~Wu, A.~D. Ronch, J.~Xiang,
  \href{http://www.sciencedirect.com/science/article/pii/S0376042116300057}{Energy
  harvesting by means of flow-induced vibrations on aerospace vehicles},
  Progress in Aerospace Sciences 86 (2016) 28 -- 62.
\newblock \href
  {https://doi.org/https://doi.org/10.1016/j.paerosci.2016.08.001}
  {\path{doi:https://doi.org/10.1016/j.paerosci.2016.08.001}}.
\newline\urlprefix\url{http://www.sciencedirect.com/science/article/pii/S0376042116300057}

\bibitem{mccarthy2016review}
J.~McCarthy, S.~Watkins, A.~Deivasigamani, S.~John,
  \href{http://www.sciencedirect.com/science/article/pii/S0022460X1500783X}{Fluttering
  energy harvesters in the wind: A review}, Journal of Sound and Vibration 361
  (2016) 355 -- 377.
\newblock \href {https://doi.org/https://doi.org/10.1016/j.jsv.2015.09.043}
  {\path{doi:https://doi.org/10.1016/j.jsv.2015.09.043}}.
\newline\urlprefix\url{http://www.sciencedirect.com/science/article/pii/S0022460X1500783X}

\bibitem{tang2009cantilevered}
L.~Tang, M.~P. Païdoussis, J.~Jiang,
  \href{http://www.sciencedirect.com/science/article/pii/S0022460X09004076}{Cantilevered
  flexible plates in axial flow: Energy transfer and the concept of
  flutter-mill}, Journal of Sound and Vibration 326~(1) (2009) 263 -- 276.
\newblock \href {https://doi.org/https://doi.org/10.1016/j.jsv.2009.04.041}
  {\path{doi:https://doi.org/10.1016/j.jsv.2009.04.041}}.
\newline\urlprefix\url{http://www.sciencedirect.com/science/article/pii/S0022460X09004076}

\bibitem{michelin_doare_2013}
S.~Michelin, O.~Doaré, Energy harvesting efficiency of piezoelectric flags in
  axial flows, Journal of Fluid Mechanics 714 (2013) 489–504.
\newblock \href {https://doi.org/10.1017/jfm.2012.494}
  {\path{doi:10.1017/jfm.2012.494}}.

\bibitem{shoele_mittal_2016}
K.~Shoele, R.~Mittal, Energy harvesting by flow-induced flutter in a simple
  model of an inverted piezoelectric flag, Journal of Fluid Mechanics 790
  (2016) 582–606.
\newblock \href {https://doi.org/10.1017/jfm.2016.40}
  {\path{doi:10.1017/jfm.2016.40}}.

\bibitem{xiao2014review}
Q.~Xiao, Q.~Zhu,
  \href{http://www.sciencedirect.com/science/article/pii/S0889974614000140}{A
  review on flow energy harvesters based on flapping foils}, Journal of Fluids
  and Structures 46 (2014) 174 -- 191.
\newblock \href
  {https://doi.org/https://doi.org/10.1016/j.jfluidstructs.2014.01.002}
  {\path{doi:https://doi.org/10.1016/j.jfluidstructs.2014.01.002}}.
\newline\urlprefix\url{http://www.sciencedirect.com/science/article/pii/S0889974614000140}

\bibitem{young2014review}
J.~Young, J.~Lai, M.~F. Platzer, A review of progress and challenges in
  flapping foil power generation, Progress in Aerospace Sciences 67 (2014)
  2--28.
\newblock \href {https://doi.org/10.1016/j.paerosci.2013.11.001}
  {\path{doi:10.1016/j.paerosci.2013.11.001}}.

\bibitem{peng2009}
Z.~Peng, Q.~Zhu, \href{https://doi.org/10.1063/1.3275852}{Energy harvesting
  through flow-induced oscillations of a foil}, Physics of Fluids 21~(12)
  (2009) 123602.
\newblock \href {http://arxiv.org/abs/https://doi.org/10.1063/1.3275852}
  {\path{arXiv:https://doi.org/10.1063/1.3275852}}, \href
  {https://doi.org/10.1063/1.3275852} {\path{doi:10.1063/1.3275852}}.
\newline\urlprefix\url{https://doi.org/10.1063/1.3275852}

\bibitem{zhu2012shear}
Q.~Zhu,
  \href{http://www.sciencedirect.com/science/article/pii/S088997461200117X}{Energy
  harvesting by a purely passive flapping foil from shear flows}, Journal of
  Fluids and Structures 34 (2012) 157 -- 169.
\newblock \href
  {https://doi.org/https://doi.org/10.1016/j.jfluidstructs.2012.05.013}
  {\path{doi:https://doi.org/10.1016/j.jfluidstructs.2012.05.013}}.
\newline\urlprefix\url{http://www.sciencedirect.com/science/article/pii/S088997461200117X}

\bibitem{wang2017structural}
Z.~Wang, L.~Du, J.~Zhao, X.~Sun,
  \href{http://www.sciencedirect.com/science/article/pii/S0889974617300312}{Structural
  response and energy extraction of a fully passive flapping foil}, Journal of
  Fluids and Structures 72 (2017) 96 -- 113.
\newblock \href
  {https://doi.org/https://doi.org/10.1016/j.jfluidstructs.2017.05.002}
  {\path{doi:https://doi.org/10.1016/j.jfluidstructs.2017.05.002}}.
\newline\urlprefix\url{http://www.sciencedirect.com/science/article/pii/S0889974617300312}

\bibitem{young2013numerical}
J.~Young, M.~A. Ashraf, J.~C. Lai, M.~F. Platzer, Numerical simulation of fully
  passive flapping foil power generation, AIAA journal 51~(11) (2013)
  2727--2739.

\bibitem{veilleux2017numerical}
J.-C. Veilleux, G.~Dumas,
  \href{http://www.sciencedirect.com/science/article/pii/S0889974616303917}{{Numerical
  optimization of a fully-passive flapping-airfoil turbine}}, Journal of Fluids
  and Structures 70 (2017) 102 -- 130.
\newblock \href
  {https://doi.org/https://doi.org/10.1016/j.jfluidstructs.2017.01.019}
  {\path{doi:https://doi.org/10.1016/j.jfluidstructs.2017.01.019}}.
\newline\urlprefix\url{http://www.sciencedirect.com/science/article/pii/S0889974616303917}

\bibitem{ramesh2015intermittent}
K.~Ramesh, J.~Murua, A.~Gopalarathnam,
  \href{http://www.sciencedirect.com/science/article/pii/S0889974615000468}{Limit-cycle
  oscillations in unsteady flows dominated by intermittent leading-edge vortex
  shedding}, Journal of Fluids and Structures 55 (2015) 84 -- 105.
\newblock \href
  {https://doi.org/https://doi.org/10.1016/j.jfluidstructs.2015.02.005}
  {\path{doi:https://doi.org/10.1016/j.jfluidstructs.2015.02.005}}.
\newline\urlprefix\url{http://www.sciencedirect.com/science/article/pii/S0889974615000468}

\bibitem{wang2018lco}
E.~Wang, K.~Ramesh, S.~Killen, I.~M. Viola,
  \href{http://www.sciencedirect.com/science/article/pii/S0889974618303475}{On
  the nonlinear dynamics of self-sustained limit-cycle oscillations in a
  flapping-foil energy harvester}, Journal of Fluids and Structures 83 (2018)
  339 -- 357.
\newblock \href
  {https://doi.org/https://doi.org/10.1016/j.jfluidstructs.2018.09.005}
  {\path{doi:https://doi.org/10.1016/j.jfluidstructs.2018.09.005}}.
\newline\urlprefix\url{http://www.sciencedirect.com/science/article/pii/S0889974618303475}

\bibitem{boudreau2018}
M.~Boudreau, G.~Dumas, M.~Rahimpour, P.~Oshkai,
  \href{http://www.sciencedirect.com/science/article/pii/S0889974618302287}{{Experimental
  investigation of the energy extraction by a fully-passive flapping-foil
  hydrokinetic turbine prototype}}, Journal of Fluids and Structures 82 (2018)
  446 -- 472.
\newblock \href
  {https://doi.org/https://doi.org/10.1016/j.jfluidstructs.2018.07.014}
  {\path{doi:https://doi.org/10.1016/j.jfluidstructs.2018.07.014}}.
\newline\urlprefix\url{http://www.sciencedirect.com/science/article/pii/S0889974618302287}

\bibitem{pigolotti2017destabilizing}
L.~Pigolotti, C.~Mannini, G.~Bartoli,
  \href{https://doi.org/10.1007/s11012-016-0604-y}{Destabilizing effect of
  damping on the post-critical flutter oscillations of flat plates}, Meccanica
  52~(13) (2017) 3149--3164.
\newblock \href {https://doi.org/10.1007/s11012-016-0604-y}
  {\path{doi:10.1007/s11012-016-0604-y}}.
\newline\urlprefix\url{https://doi.org/10.1007/s11012-016-0604-y}

\bibitem{pigolotti2017jsv}
L.~Pigolotti, C.~Mannini, G.~Bartoli, K.~Thiele,
  \href{http://www.sciencedirect.com/science/article/pii/S0022460X17304030}{Critical
  and post-critical behaviour of two-degree-of-freedom flutter-based
  generators}, Journal of Sound and Vibration 404 (2017) 116 -- 140.
\newblock \href {https://doi.org/https://doi.org/10.1016/j.jsv.2017.05.024}
  {\path{doi:https://doi.org/10.1016/j.jsv.2017.05.024}}.
\newline\urlprefix\url{http://www.sciencedirect.com/science/article/pii/S0022460X17304030}

\bibitem{pigolotti2017jfs}
L.~Pigolotti, C.~Mannini, G.~Bartoli,
  \href{http://www.sciencedirect.com/science/article/pii/S0889974617300786}{Experimental
  study on the flutter-induced motion of two-degree-of-freedom plates}, Journal
  of Fluids and Structures 75 (2017) 77 -- 98.
\newblock \href
  {https://doi.org/https://doi.org/10.1016/j.jfluidstructs.2017.07.014}
  {\path{doi:https://doi.org/10.1016/j.jfluidstructs.2017.07.014}}.
\newline\urlprefix\url{http://www.sciencedirect.com/science/article/pii/S0889974617300786}

\bibitem{bryant2012wake}
M.~Bryant, R.~L. Mahtani, E.~Garcia, Wake synergies enhance performance in
  aeroelastic vibration energy harvesting, Journal of Intelligent Material
  Systems and Structures 23~(10) (2012) 1131--1141.
\newblock \href {https://doi.org/10.1177/1045389X12443599}
  {\path{doi:10.1177/1045389X12443599}}.

\bibitem{kirschmeier2018wake}
B.~Kirschmeier, M.~Bryant,
  \href{http://www.sciencedirect.com/science/article/pii/S0889974617306242}{Experimental
  investigation of wake-induced aeroelastic limit cycle oscillations in tandem
  wings}, Journal of Fluids and Structures 81 (2018) 309 -- 324.
\newblock \href
  {https://doi.org/https://doi.org/10.1016/j.jfluidstructs.2018.04.015}
  {\path{doi:https://doi.org/10.1016/j.jfluidstructs.2018.04.015}}.
\newline\urlprefix\url{http://www.sciencedirect.com/science/article/pii/S0889974617306242}

\bibitem{mccarthy2013downstream}
J.~McCarthy, A.~Deivasigamani, S.~John, S.~Watkins, F.~Coman, P.~Petersen,
  \href{http://www.sciencedirect.com/science/article/pii/S0894177713001854}{Downstream
  flow structures of a fluttering piezoelectric energy harvester}, Experimental
  Thermal and Fluid Science 51 (2013) 279 -- 290.
\newblock \href
  {https://doi.org/https://doi.org/10.1016/j.expthermflusci.2013.08.010}
  {\path{doi:https://doi.org/10.1016/j.expthermflusci.2013.08.010}}.
\newline\urlprefix\url{http://www.sciencedirect.com/science/article/pii/S0894177713001854}

\bibitem{mccarthy2014visualisation}
J.~McCarthy, A.~Deivasigamani, S.~Watkins, S.~John, F.~Coman, P.~Petersen,
  \href{http://www.sciencedirect.com/science/article/pii/S0894177714001381}{On
  the visualisation of flow structures downstream of fluttering piezoelectric
  energy harvesters in a tandem configuration}, Experimental Thermal and Fluid
  Science 57 (2014) 407 -- 419.
\newblock \href
  {https://doi.org/https://doi.org/10.1016/j.expthermflusci.2014.05.017}
  {\path{doi:https://doi.org/10.1016/j.expthermflusci.2014.05.017}}.
\newline\urlprefix\url{http://www.sciencedirect.com/science/article/pii/S0894177714001381}

\bibitem{zhang2000flexible}
J.~Zhang, S.~Childress, A.~Libchaber, M.~Shelley,
  \href{http://dx.doi.org/10.1038/35048530}{Flexible filaments in a flowing
  soap film as a model for one-dimensional flags in a two-dimensional wind},
  Nature 408~(6814) (2000) 835.
\newblock \href {https://doi.org/10.1038/35048530}
  {\path{doi:10.1038/35048530}}.
\newline\urlprefix\url{http://dx.doi.org/10.1038/35048530}

\bibitem{favier2015}
J.~Favier, A.~Revell, A.~Pinelli,
  \href{http://www.sciencedirect.com/science/article/pii/S0889974614002643}{Numerical
  study of flapping filaments in a uniform fluid flow}, Journal of Fluids and
  Structures 53 (2015) 26 -- 35, special Issue on Unsteady Separation in
  Fluid-Structure Interaction–II.
\newblock \href
  {https://doi.org/https://doi.org/10.1016/j.jfluidstructs.2014.11.010}
  {\path{doi:https://doi.org/10.1016/j.jfluidstructs.2014.11.010}}.
\newline\urlprefix\url{http://www.sciencedirect.com/science/article/pii/S0889974614002643}

\bibitem{huertas-cerdeira2018}
C.~Huertas-Cerdeira, B.~Fan, M.~Gharib,
  \href{http://www.sciencedirect.com/science/article/pii/S0889974617303973}{Coupled
  motion of two side-by-side inverted flags}, Journal of Fluids and Structures
  76 (2018) 527 -- 535.
\newblock \href
  {https://doi.org/https://doi.org/10.1016/j.jfluidstructs.2017.11.005}
  {\path{doi:https://doi.org/10.1016/j.jfluidstructs.2017.11.005}}.
\newline\urlprefix\url{http://www.sciencedirect.com/science/article/pii/S0889974617303973}

\bibitem{boragno2012elastically}
C.~Boragno, R.~Festa, A.~Mazzino,
  \href{https://doi.org/10.1063/1.4729936}{Elastically bounded flapping wing
  for energy harvesting}, Applied Physics Letters 100~(25) (2012) 253906.
\newblock \href {http://arxiv.org/abs/https://doi.org/10.1063/1.4729936}
  {\path{arXiv:https://doi.org/10.1063/1.4729936}}, \href
  {https://doi.org/10.1063/1.4729936} {\path{doi:10.1063/1.4729936}}.
\newline\urlprefix\url{https://doi.org/10.1063/1.4729936}

\bibitem{orchini2013flapping}
A.~Orchini, A.~Mazzino, J.~Guerrero, R.~Festa, C.~Boragno,
  \href{https://doi.org/10.1063/1.4821808}{Flapping states of an elastically
  anchored plate in a uniform flow with applications to energy harvesting by
  fluid-structure interaction}, Physics of Fluids 25~(9) (2013) 097105.
\newblock \href {http://arxiv.org/abs/https://doi.org/10.1063/1.4821808}
  {\path{arXiv:https://doi.org/10.1063/1.4821808}}, \href
  {https://doi.org/10.1063/1.4821808} {\path{doi:10.1063/1.4821808}}.
\newline\urlprefix\url{https://doi.org/10.1063/1.4821808}

\bibitem{olivieri2017fluttering}
S.~Olivieri, G.~Boccalero, A.~Mazzino, C.~Boragno,
  \href{http://www.sciencedirect.com/science/article/pii/S0960148116311260}{Fluttering
  conditions of an energy harvester for autonomous powering}, Renewable Energy
  105 (2017) 530 -- 538.
\newblock \href {https://doi.org/https://doi.org/10.1016/j.renene.2016.12.067}
  {\path{doi:https://doi.org/10.1016/j.renene.2016.12.067}}.
\newline\urlprefix\url{http://www.sciencedirect.com/science/article/pii/S0960148116311260}

\bibitem{olivieri2017aeroelastic}
S.~Olivieri, G.~Boccalero, A.~Mazzino, C.~Boragno,
  \href{http://www.sciencedirect.com/science/article/pii/S1877705817339504}{{FLuttering
  Energy Harvester for Autonomous Powering (FLEHAP): aeroelastic
  characterisation and preliminary performance evaluation}}, Procedia
  Engineering 199 (2017) 3474 -- 3479, {X International Conference on
  Structural Dynamics, EURODYN 2017}.
\newblock \href {https://doi.org/https://doi.org/10.1016/j.proeng.2017.09.456}
  {\path{doi:https://doi.org/10.1016/j.proeng.2017.09.456}}.
\newline\urlprefix\url{http://www.sciencedirect.com/science/article/pii/S1877705817339504}

\bibitem{boccalero2017power}
G.~Boccalero, S.~Olivieri, A.~Mazzino, C.~Boragno,
  \href{http://stacks.iop.org/0964-1726/26/i=9/a=095031}{Power harvesting by
  electromagnetic coupling from wind-induced limit cycle oscillations}, Smart
  Materials and Structures 26~(9) (2017) 095031.
\newline\urlprefix\url{http://stacks.iop.org/0964-1726/26/i=9/a=095031}

\bibitem{ferziger2012computational}
J.~H. Ferziger, M.~Peric, Computational methods for fluid dynamics, Springer
  Science \& Business Media, 2012.

\bibitem{mittal2005_review}
R.~Mittal, G.~Iaccarino,
  \href{https://doi.org/10.1146/annurev.fluid.37.061903.175743}{Immersed
  boundary methods}, Annual Review of Fluid Mechanics 37~(1) (2005) 239--261.
\newblock \href
  {http://arxiv.org/abs/https://doi.org/10.1146/annurev.fluid.37.061903.175743}
  {\path{arXiv:https://doi.org/10.1146/annurev.fluid.37.061903.175743}}, \href
  {https://doi.org/10.1146/annurev.fluid.37.061903.175743}
  {\path{doi:10.1146/annurev.fluid.37.061903.175743}}.
\newline\urlprefix\url{https://doi.org/10.1146/annurev.fluid.37.061903.175743}

\bibitem{detullio2016}
M.~de~Tullio, G.~Pascazio,
  \href{http://www.sciencedirect.com/science/article/pii/S0021999116303692}{A
  moving-least-squares immersed boundary method for simulating the
  fluid-structure interaction of elastic bodies with arbitrary thickness},
  Journal of Computational Physics 325 (2016) 201 -- 225.
\newblock \href {https://doi.org/https://doi.org/10.1016/j.jcp.2016.08.020}
  {\path{doi:https://doi.org/10.1016/j.jcp.2016.08.020}}.
\newline\urlprefix\url{http://www.sciencedirect.com/science/article/pii/S0021999116303692}

\bibitem{verzicco1996}
R.~Verzicco, P.~Orlandi,
  \href{http://www.sciencedirect.com/science/article/pii/S0021999196900339}{A
  finite-difference scheme for three-dimensional incompressible flows in
  cylindrical coordinates}, Journal of Computational Physics 123~(2) (1996) 402
  -- 414.
\newblock \href {https://doi.org/https://doi.org/10.1006/jcph.1996.0033}
  {\path{doi:https://doi.org/10.1006/jcph.1996.0033}}.
\newline\urlprefix\url{http://www.sciencedirect.com/science/article/pii/S0021999196900339}

\bibitem{spandan2017}
V.~Spandan, V.~Meschini, R.~Ostilla-Mónico, D.~Lohse, G.~Querzoli, M.~D.
  de~Tullio, R.~Verzicco,
  \href{http://www.sciencedirect.com/science/article/pii/S0021999117305442}{A
  parallel interaction potential approach coupled with the immersed boundary
  method for fully resolved simulations of deformable interfaces and
  membranes}, Journal of Computational Physics 348 (2017) 567 -- 590.
\newblock \href {https://doi.org/https://doi.org/10.1016/j.jcp.2017.07.036}
  {\path{doi:https://doi.org/10.1016/j.jcp.2017.07.036}}.
\newline\urlprefix\url{http://www.sciencedirect.com/science/article/pii/S0021999117305442}

\bibitem{vanella2009}
M.~Vanella, E.~Balaras,
  \href{http://www.sciencedirect.com/science/article/pii/S0021999109003246}{A
  moving-least-squares reconstruction for embedded-boundary formulations},
  Journal of Computational Physics 228~(18) (2009) 6617 -- 6628.
\newblock \href {https://doi.org/https://doi.org/10.1016/j.jcp.2009.06.003}
  {\path{doi:https://doi.org/10.1016/j.jcp.2009.06.003}}.
\newline\urlprefix\url{http://www.sciencedirect.com/science/article/pii/S0021999109003246}

\bibitem{of2018}
\href{http://www.openfoam.com}{{OpenFOAM}, {The} open source {CFD} toolbox}.
\newline\urlprefix\url{http://www.openfoam.com}

\end{thebibliography}
%

%
%
%

\end{document}